\documentclass[articlepaper,twocolumn,aps,prapplied,amsmath,amssymb,superscriptaddress,floats,floatfix]{revtex4-2}
\usepackage[dvipsnames]{xcolor}
\usepackage{graphicx}
\usepackage{dcolumn}
\usepackage{bm}
\usepackage{amsfonts}
\usepackage{graphicx}
\usepackage{threeparttable}
\usepackage{multirow}
\usepackage{array}
\usepackage{float}
\usepackage{soul}
\usepackage[colorlinks, citecolor=red]{hyperref}
\usepackage{amsthm,amsmath,amssymb}
\usepackage{mathrsfs}
\newcommand{\ket}[1]{\vert #1 \rangle}
\newcommand{\bra}[1]{\langle #1 \vert}

\begin{document}


\title{Transmon-assisted high-fidelity controlled-Z gates for integer fluxonium qubits}
\author{J.-H. Wang}
\thanks{These two authors contributed equally to this work.}
\affiliation{Center for Quantum Information, Institute for Interdisciplinary Information Sciences, Tsinghua University, Beijing 100084, China}
\author{H. Xiong}
\thanks{These two authors contributed equally to this work.}
\affiliation{Center for Quantum Information, Institute for Interdisciplinary Information Sciences, Tsinghua University, Beijing 100084, China}
\affiliation{Current address: Beijing Key Laboratory of Fault-Tolerant Quantum Computing, Beijing Academy of Quantum Information Sciences, Beijing 100193, China}
\author{J.-Z. Yang}
\affiliation{Center for Quantum Information, Institute for Interdisciplinary Information Sciences, Tsinghua University, Beijing 100084, China}
\author{H.-Y. Zhang}
\affiliation{Center for Quantum Information, Institute for Interdisciplinary Information Sciences, Tsinghua University, Beijing 100084, China}
\affiliation{Hefei National Laboratory, Hefei 230088, China}

\author{Y.-P. Song}\email{ypsong@mail.tsinghua.edu.cn}
\affiliation{Center for Quantum Information, Institute for Interdisciplinary Information Sciences, Tsinghua University, Beijing 100084, China}
\affiliation{Hefei National Laboratory, Hefei 230088, China}

\author{L.-M. Duan}\email{lmduan@tsinghua.edu.cn}
\affiliation{Center for Quantum Information, Institute for Interdisciplinary Information Sciences, Tsinghua University, Beijing 100084, China}
\affiliation{Hefei National Laboratory, Hefei 230088, China}





\begin{abstract}
Fluxoniums, as partially-protected superconducting qubits are promising to be employed to build high-performance large-scale quantum processor. The recently proposed ``integer fluxonium" operates at zero external flux bias, with a frequency of approximately 3 GHz. Single-qubit gate fidelity has been demonstrated to exceed $99.9\%$ \cite{mencia2024integerfluxoniumqubit}, while two-qubit gate schemes and scalable architectures remain underexplored. In this work, we investigate a fluxonium-transmon-fluxonium (FTF) coupling architecture using integer fluxoniums. We first confirm suppression of $ZZ$ interaction in the FTF system and then propose two high-fidelity controlled-$Z$ (CZ) gate schemes utilizing the coupler control: a flux-activated adiabatic gate scheme and a microwave-activated non-adiabatic gate scheme. Both schemes are capable of achieving low coherent error on the order of $1 \times 10^{-6}$ within gate durations of several tens of nanoseconds. Additionally, we discuss a hybrid circuit system in which an integer fluxonium is coupled to a conventional fluxonium through a transmon coupler. Our proposal provides insights for future implementations of large-scale quantum circuits based on integer fluxonium devices.

\end{abstract}


\maketitle

\section{INTRODUCTION}\label{Sec1}
Despite the success of transmon qubits in state-of-the-art superconducting processors, progressing beyond the NISQ era demands novel noise-resistant qubit designs \cite{PRXQuantum.2.010339,Pechenezhskiy2020TheSQ,Chou2024ASD,PhysRevLett.132.180601,Ni2022BeatingTB} and innovative coupling architectures \cite{PhysRevApplied.10.054062,emma,PhysRevResearch.4.043141,PhysRevLett.113.220502,PhysRevLett.119.180511,ring,PhysRevLett.127.080505,PRXQuantum.4.040321}. Transmon qubits are initially designed to address charge offset noise \cite{transmon} and later become a building block of large scale superconducting circuits \cite{Arute2019QuantumSU,Acharya2024QuantumEC,ZHU2022240,PhysRevLett.127.180501,NonAbelianbraidingofFibonaccianyons,Zhang2023} after more than a decade of improvement. The simple structure of transmon is convenient for design and fabrication but also brings disadvantages such as susceptibility to dielectric noise, weak anharmonicity and inflexible spectrum. These problems can be reduced by introducing a large inductance to the circuit, namely by replacing the transmon qubit with the fluxonium qubit \cite{PhysRevX.9.041041,PhysRevLett.103.217004,Wang2022,PRXQuantum.3.037001,PhysRevLett.130.267001,wang2024achievingmillisecondcoherencefluxonium}. Compared to transmon, fluxonium is a type of less exploited qubit with increasing popularity. In recent years, a variety of entangling gates have been demonstrated in fluxonium systems \cite{ld,PRXQuantum.5.020326,Moskalenko2022,PhysRevResearch.4.023040,PhysRevX.11.021026,PhysRevApplied.20.024011,PhysRevLett.132.060602,emma,PRXQuantum.4.040321}, revealing their potential for large scale applications. A conventional fluxonium operates at half-flux quanta with a typical qubit frequency below 1 GHz. When the fluxonium parameters are designed such that the fluxon transition is much lower than the plasmon transition, namely $\sqrt{8E_JE_C}\gg2\pi^2E_L$, the circuit enters the so-called integer fluxonium (IF) regime where the qubits are also protected at zero flux quanta with much higher frequency \cite{ardati2024usingbifluxontunnelingprotect,mencia2024integerfluxoniumqubit}. Such an option greatly extends the available frequency range and allows more diverse spectrum design. On the other hand, the ability to work at zero flux quanta also releases the requirements for flux tunability. Integrating this new type of qubit into large-scale circuits may bring new opportunities to the field. Therefore, it is desirable to construct a scalable entangling scheme with high-fidelity two-qubit gates on integer fluxonium qubits to unleash their full potential.

In this work, we consider a transmon-coupler-assisted integer-fluxonium system. When two integer fluxoniums are coupled by a tunable transmon, the static $ZZ$ interaction can be completely eliminated at proper system parameters without confronting the straddling regime issue in the all-transmon architecture \cite{PhysRevApplied.10.054062,PhysRevX.11.021058,PhysRevApplied.12.054023,PhysRevLett.125.240503,PhysRevLett.127.080505,PhysRevLett.125.240502,DiCarlo,PhysRevA.90.022307,PhysRevApplied.14.024070,PhysRevApplied.16.024037,fors2024comprehensiveexplanationzzcoupling}. The absence of $ZZ$ interaction originates from the cancellation effect from the qubit direct coupling and the coupler-mediated coupling, along with the unique spectrum of integer fluxonium. Similar to the case of transmon systems, a flux-activated CZ gate can be realized by sending an adiabatic pulse to the coupler flux line. Due to the large anharmonicity of the integer fluxonium, the $\ket{2}$ states are far detuned from other states in the double-excitation manifold of the coupled system. Therefore the harmful small-gap anti-crossings can be avoided in the adiabatic trajectory  \cite{PhysRevApplied.16.054020}. To suppress the leakage error during the flux pulse, we design a pulse edge to ensure a constant transition rate and then optimize the pulse plateau to further suppress the leakage error down to $10^{-6}$ level.  We then incorporate energy relaxation and $1/f$ flux noise \cite{10.1063/1.98041,RevModPhys.86.361,PhysRevLett.130.220602}
 as the source of incoherent errors in error budget estimation. The total error can reach a level of $10^{-4}$ for typical qubits' $T_1$ and a flux noise intensity of $A_{\varphi}=1\times 10^{-6}{\rm Hz}^{-\frac{1}{2}}$.
 
 An alternative gate scheme is to employ $2\pi$ rotation on the coupler transition with a microwave drive to acquire a geometric conditional phase. We utilize the level repulsion between the coupler's first excited state and the third excited state of the fluxoniums. The microwave pulse is delivered through the coupler's dedicated charge line. Compared with Ref.~\cite{ld}, our approach avoids simultaneous drives on both fluxoniums, which suppresses crosstalk and simplifies the requirements on the control. By optimizing the circuit parameters, the leakage error can be reduced to $10^{-5}$ with a 70 ns duration Gaussian pulse. Besides the target transition, other off-resonant transitions can also contribute to the conditional phase through ac-Stark shifts. These unwanted phase accumulations can be compensated by introducing a detuning to the microwave drive. We also investigate the gate fidelity in the presence of dissipation, and the results indicate that achieving a CZ gate fidelity exceeding 99.99\% is feasible when the qubits’ $T_1$ exceeds 500 µs and the target states’ $T_1$ exceeds 50 $\mu s$.

In addition, we discuss a hybrid circuit system, in which an integer fluxonium is coupled to a conventional fluxonium through a transmon coupler. The large qubit detuning ensures suppressing single-qubit drive crosstalk. Similar gate schemes can be implemented, but one needs to carefully re-design the system spectrum and choose appropriate transitions involved in the gate.

The paper is organized as follows. In Sec.~\ref{sec:FTF}, we calculate the static $ZZ$ interaction of the fluxonium-transmon-fluxonium (FTF) Hamiltonian. In Sec.~\ref{sec:fluxCZ}, we introduce the CZ gate scheme by adiabaticaly tuning the coupler flux and then discuss the microwave-activated gate in Sec.~\ref{sec:microwaveCZ}. After that we consider the hybrid system of integer fluxonium and conventional fluxonium in Sec.~\ref{sec:IFTF}. Finally we summarize the advantage of our proposal and discuss the incoherence error and the scalability of the scheme. 

\section{Static ZZ interaction\label{sec:FTF}}
\begin{figure} 
    \includegraphics[width=0.9\linewidth,keepaspectratio]{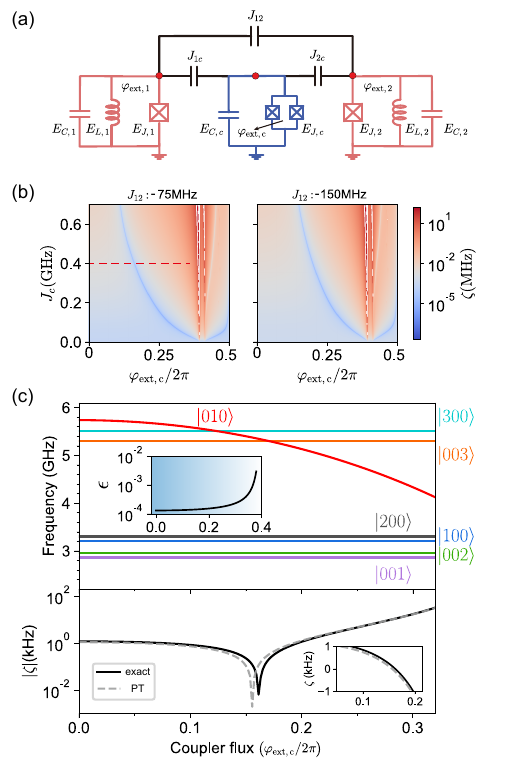}
    \caption{\label{fig:test1}FTF system circuit and $ZZ$ interaction. 
    (a) Circuit schematic of FTF Hamiltonian. 
    (b) Numerically calculated $\zeta$ as a function of $J_c$ and $\varphi_{\rm ext,c}$. $J_{12}=-75$ MHz in left panel and $J_{12}=-150$ MHz in right panel. The white pixels represent that at least one computational state is highly hybridized such that no well defined dressed state corresponding to the bare state.
    (c) The upper panel shows bare frequencies of the coupler $\ket{1}$ state and the fluxoniums $\ket{1}$, $\ket{2}$, $\ket{3}$ states. Inset shows the delocalization parameter which is defined as $\epsilon=\max(|\bra{\widetilde{100}}001\rangle|^2,|\bra{\widetilde{001}}100\rangle|^2)$. The bottom panel plots $\zeta$ as a function of $\varphi_{\rm ext,c}$, where the black solid curve shows the exact result from numerical diagonalization, while the dashed curve is simulated by perturbation theory. The parameters used in (c) is marked by the red dashed line in (b).}
    \end{figure}
Here we discuss the FTF architecture, where two fluxonium qubits are capacitively coupled through a transmon coupler. The Hamiltonian of this system can be described as:
\begin{equation}\label{eq:ftf1}
    \begin{aligned}
        H/h = &\sum_{i=1,2}[4E_{C,i}\hat{n}_i^2+\frac{1}{2}E_{L,i}\hat{\varphi}_i^2
        -E_{J,i} \cos(\hat{\varphi}_i-\varphi_{\rm ext,i})]\\
        +& 4E_{C,c}\hat{n}_c^2-E_{J,c} \cos({\varphi_{\rm ext,c}})\cos(\hat{\varphi}_c)\\
        +&J_{12}\hat{n}_1\hat{n}_2+J_{1c}\hat{n}_1\hat{n}_{c}+J_{2c}\hat{n}_2\hat{n}_{c},
    \end{aligned}
\end{equation}
where $i = 1, 2$ index the two fluxonium modes, and $c$ labels the transmon coupler mode. $E_c$, $E_J$ and $E_L$ represent charging energy, Josephson energy and inductive energy respectively. $\varphi_{\rm ext}$ is the external flux bias.  The conventional fluxonium is working at the half-flux quanta ($\varphi_{\rm ext}=0.5\times 2\pi$), while integer fluxonium is working at integer flux quanta ($\varphi_{\rm ext}=0$). In this paper we describe the quantum state of the uncoupled system using the notation $\ket{ijk}$, where $i$, $j$, and $k$ denote the energy eigenstates in the basis of fluxonium 1, the coupler, and fluxonium 2, respectively. We use the state notation $\ket{\widetilde{ijk}}$ to represent the corresponding dressed states. For simplicity, we use the two-qubit state $\ket{ij}$ to denote $\ket{\widetilde{i0j}}$ when the coupler is in the ground state.
 
The direct and the coupler-mediated couplings between two fluxonium qubits generate a static $ZZ$ interaction strength $\zeta=(E_{11}+E_{00}-E_{10}-E_{01})/h$, where $E_{ij}$ ($i,j=0,1$) is the energy of the computational state $\ket{ij}$. This always-on entangling interaction is deleterious during the qubit idling period, and degrades the gate performance. One key feature of the FTF architecture is its ability to suppress $\zeta$. It is analytically and experimentally demonstrated in Ref.~\cite{ld} with conventional fluxoniums. Here we focus on the integer fluxonium case.

Given that a typical integer fluxonium works with a frequency range $2\sim4$ GHz \cite{ardati2024usingbifluxontunnelingprotect,mencia2024integerfluxoniumqubit}, we set the transmon coupler's frequency above the fluxoniums' frequency. Here we note that in IF-T-IF system, the proper choice of parameters enables elimination of the static $ZZ$ coupling, while the computational states are only slightly hybridized.

The static $ZZ$ coupling derives from the virtual interactions of both computational and high energy states, which can be understood with perturbation theory. By expanding the energy corrections to the fourth order, we can rewrite the $ZZ$ coupling strength as 
\begin{equation}
    \zeta=\zeta^{(2)}+\zeta^{(3)}+\zeta^{(4)},
\end{equation}
where $\zeta^{(2)}$, $\zeta^{(3)}$, $\zeta^{(4)}$ represent the second-, third-, and fourth-order corrections, respectively.
The second order correction derives from the direct coupling of the two fluxoniums with the relation $\zeta^{(2)}\propto |J_{12}|^2$, while  the third- and the fourth-order corrections are induced by coupler-mediated couplings $\zeta^{(3)}\propto J_{12}|J_{c}|^2$, $\zeta^{(4)}\propto |J_{c}|^4$(see Appendix.~\ref{ptzz} for details). Here we assume $J_c=J_{1c}=J_{2c}$ for simplicity. These relations inspire us to design proper coupling strengths and coupler frequency for destructive interference of $\zeta^{(2)}$, $\zeta^{(3)}$ and $\zeta^{(4)}$. We note that a grounded transmon coupler gives a relative constant coupling ratio as $J_{c}/J_{12}\sim 5$, while using a differential transmon coupler will lead to an opposite sign between $J_{12}$ and $J_c$, and a tunable ratio $|J_{c}/J_{12}|$ (see Appendix.~\ref{apd:a} for details). 

Fig.~\ref{fig:test1}(b) and (c) illustrate the full cancellation of $ZZ$ coupling in a IF-T-IF system with qubit parameters: $E_{C,1(2)}=1.6$ GHz, $E_{J,1(2)}=4.1(3.9)$ GHz, $E_{L,1(2)}=0.18(0.16)$ GHz, $E_{C,c}=250$ MHz, $E_{J,c}=18$ GHz. 
Fig.~\ref{fig:test1}(b) shows the $ZZ$ cancellation point exists for various coupling strengths. The white pixels represent at least one computational state in this region is highly hybridized. Here a highly hybridized state refers to the state that has an overlap $<0.5$ with all the system's bare states. 

In addition to parasitic $ZZ$, one also expects the delocalization effect to be suppressed, which can be characterized by a delocalization parameter \cite{PhysRevApplied.20.064037}:
\begin{equation}
    \epsilon=\max(|\bra{\widetilde{100}}001\rangle|^2,|\bra{\widetilde{001}}100\rangle|^2).
\end{equation}
 The delocalization parameter measures excitation exchange between idle qubits. This parameter captures a distinct class of unwanted couplings that complement the well-characterized $ZZ$ interactions, providing a more complete picture of inter-qubit crosstalk. A large delocalization parameter $\epsilon$ indicates stronge inter-qubit coupling, which induces significant quantum crosstalk. This effect represents one of the primary challenges in scaling up superconducting quantum processors. By numerically evaluating these parameters, we demonstrate our coupler design simultaneously suppresses unwanted $ZZ$ interactions and maintains a delocalization at $~10^{-4}$ level as shown in Fig.~\ref{fig:test1}(c). 
\begin{figure} 
    \includegraphics[width=0.9\linewidth,keepaspectratio]{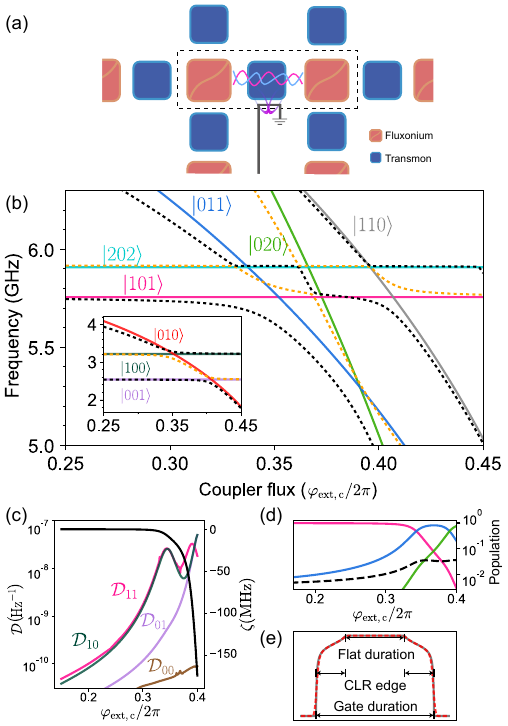}
    \caption{\label{fig:fig2}Adiabatic controlled-phase gate simulation with $E_{C,1(2)}=1.5$ GHz, $E_{J,1(2)}=4.1(3.8)$ GHz, $E_{L,1(2)}=0.18(0.14)$ GHz, $E_{C,c}=180$ MHz, $E_{J,c}=18$ GHz, $J_{c,1(2)}=600$ MHz, and $J_{12}=-100$ MHz. (a) Sketch of the adiabatic CZ gate architecture. The flux pulse is delivered via the couplers' dedicated flux line to the FTF system. (b) Frequencies of the bare states (solid lines) and eigenstates (dashed black and orange lines) in the even-excitation manifold versus coupler flux $\varphi_{\rm ext,c}$. Inset: frequencies of the states in the single-excitation manifold. (c) Calculated leakage rate factor $ {\mathcal D}$ at various coupler flux $\varphi_{\rm ext,c}$. The black curve represents the simulated ZZ interaction as a function of the coupler flux. (d) State overlap between the bare states and the dressed state $\widetilde{\ket{101}}$ at various $\varphi_{\rm ext,c}$. The pink, blue, and green solid lines are $|\bra{\widetilde{101}}101\rangle|^2$,$|\bra{\widetilde{101}}011\rangle|^2$ and $|\bra{\widetilde{101}}020\rangle|^2$ respectively. The dashed black line shows the total probability of other bare states, defined as $1-|\bra{\widetilde{101}}101\rangle|^2-|\bra{\widetilde{101}}011\rangle|^2-|\bra{\widetilde{101}}020\rangle|^2$. (e) Sketch of the flux pulse. The pulse is composed of the constant-leakage-rate edge (CLR edge) and the flat duration. The dashed red line represents the actual pulse in our simulation, which is obtained by adding extra 10 ns idle duration to the original pulse.} 
\end{figure} 
\section{ADIABATIC CZ GATE\label{sec:fluxCZ}} 
\subsection{Gate principle}
 By tuning down the coupler frequency from the idle frequency, we can obtain strong $ZZ$ interaction in the region where the coupler is near-resonant with the fluxonium. In this regime, a $ZZ$ coupling $|\zeta|$ at the level of 100 MHz primarily arises from the different frequency shifts of the computational states $\ket{\widetilde{101}}$ and $\ket{\widetilde{100}}$. One is induced by the $\ket{101}-\ket{011}-\ket{020}$ interaction in the double-excitation manifold, and the other is from the $\ket{100}-\ket{010}-\ket{001}$ interaction in the single-excitation manifold. Without loss of generality, we use the first index to label the fluxonium with higher frequency below and hereafter. The gate process can be described as follows. When the coupler frequency goes down, $\ket{\widetilde{101}}$ is first pushed by $\ket{011}$, while $|\zeta|$ is not changed by much, because $\ket{\widetilde{100}}$ is also pushed by $\ket{010}$ with the same strength ($J_c n_{c,01}n_{f,01}$). Passing through the avoided-crossing point, the maximal overlapped bare state of $\ket{\widetilde{101}}$ changes from $\ket{101}$ to $\ket{011}$ (as shown in Fig.~\ref{fig:fig2}(d)). As $\ket{011}$ further goes down, $|\zeta|$ begins to rise drastically due to the interaction between $\ket{020}$ and $\ket{011}$. At the same time, the maximal overlapped bare state of $\ket{\widetilde{101}}$ now changes from $\ket{011}$ to $\ket{020}$.

The gate process can be described as an effective Hamiltonian:
\begin{equation}
    H_{\rm eff}/h = \chi_{00}\ket{\widetilde{000}}+\chi_{01}\ket{\widetilde{001}}+\chi_{10}\ket{\widetilde{100}}+\chi_{11}\ket{\widetilde{101}}.
\end{equation}
The $ZZ$ interaction strength is $\zeta=\chi_{11}+\chi_{00}-\chi_{10}-\chi_{01}$, and $\chi_{ij}$ ($i,j=0,1$) is a function of $\varphi_{\rm ext,c}$ originating from the flux dependence of the coupler frequency and the charge matrix elements.

A similar gate scheme has been demonstrated and widely used in the all-transmon architecture \cite{PhysRevLett.125.240503,PhysRevApplied.16.054020,PhysRevLett.125.240502,PhysRevLett.127.080505}. Before investigating the gate performance in the FTF architecture, we list two crucial differences with regard to the adiabatic dynamics between these two architectures. 

\paragraph{Charge matrix element}
The charge matrix element is defined as $n_{m,ij}=\bra{i}\hat{n}_m\ket{j}$, where $m$ denotes the quantum object, and $i(j)$ labels its eigenstates.
Transmon's charge matrix element for $0-1$ transition is remarkably larger than that of  integer fluxonium ($n_{t,01}\sim 1$ and $n_{f,01}\sim 0.15$ in our example). So the qubit-coupler coupling strengths $J_c$ in the FTF system need to be scaled correspondingly to maintain the same energy level repulsion as that in the all-transmon architecture.

\paragraph{Energy level structure}
In all-transmon architectures, the second excited state ($\ket{2}$) lies close in energy to other states in the double-excitation manifold. Due to their weak coupling to the coupler, these states can exhibit small avoided crossings when the coupler flux is tuned. Consequently, the adiabatic condition becomes more challenging to satisfy during this process, increasing the likelihood of population leakage into the second excited states. In contrast, owing to the large anharmonicity of the integer fluxonium, the $\ket{2}$ state is energetically well-separated from other states in the double-excitation manifold of the coupled system. This strong detuning effectively eliminates problematic small-gap avoided crossings along the adiabatic trajectory, significantly reducing the risk of leakage errors during gate operations.

\subsection{\label{optad} Optimizing pulse shape}
To implement a controlled-phase gate, the total $ZZ$ phase accumulates as $\theta_{zz}=\int_{0}^{T_g}\zeta(t)=(1+2N)\pi$, where $N$ is an integer. We design the flux pulse shape as shown in Fig.~\ref{fig:fig2}(e), where the pulse is symmetric with a rising(falling) edge and a flat holding period. In simulation, we add 5 ns extra idling duration to the beginning and the end of the original pulse and smooth the entire pulse with a Gaussian filter of 2 ns.
Here we apply a two-step method to optimize the pulse for leakage suppression. In the first step, we design a constant transition rate edge. We quantify the unwanted transition rate of a computational state $\ket{\widetilde{i0j}}$ by a parameter $\beta_{ij}={\mathcal D}_{ij}|\frac{\partial \varphi_{\rm ext,c}}{\partial t}|$, and the factor ${\mathcal D}_{ij}$ is defined as \cite{PhysRevApplied.16.054020}:
\begin{equation}
    {\mathcal D}_{ij}=\sum_{n}|\bra{\widetilde{n}}{\frac{\partial H}{\partial \varphi_{\rm ext,c}}}\ket{\widetilde{i0j}}/(E_n-E_{ij})^2|/\hbar,
\end{equation}
where n labels all other eigenstates except for the computational state $\ket{ij}$. These factors represent how well the adiabatic condition is satisfied during the gate, and they are functions of coupler flux $\varphi_{\rm ext,c}$. Fig.~\ref{fig:fig2}(c) plots the calculated ${\mathcal D_{ij}}$ curves, and the peaks are caused by avoided crossings, where the unwanted transitions are more likely to occur. Note that leakage error is defined by fidelity (see Appendix.~\ref{ap:ies} for details) of a controlled phase $\pi+\delta \theta$ gate, where $\delta \theta$ is an optimized small phase deviation for the best gate fidelity. We do this to separate the phase error from the leakage error, and we emphasize that the phase deviation is reasonable because the Gaussian filter slows down the drastic variations of the pulse and results in deviations from the calculated trajectory for the exact phase of $\pi$.
The principle of our optimization strategy is to keep a constant leakage error rate during the edge so that the flux changes fast in favorable adiabatic condition and moves slowly in the opposite situation. To do so, we take the sum of ${\mathcal D_{ij}}$ to obtain $\bar{{\mathcal D}}\equiv {\mathcal D_{00}}+{\mathcal D_{01}}+{\mathcal D_{10}}+{\mathcal D_{11}}$, and ensure
\begin{equation}
|\frac{\partial\varphi_{\rm ext,c}}{\partial t}|=|1/\bar{{\mathcal D}}(\varphi_{\rm ext,c})|\beta,
\end{equation}
during the rising and falling edges. In this way, the edge shape can be calculated by solving the corresponding differential equation numerically. The flux ramping rate is generally controlled by the parameter $\beta$. Therefore, for a given edge length, $\beta$ determines the flux with the flat top, along with the largest $ZZ$ rate $|\zeta|$. Overall, the conditional phase accumulated during the pulse increases monotonically with $\beta$. As a result, one can always find an exact value of $\beta$ that yields a conditional $\pi$ phase for a given set of edge length and flat length, as long as the maximal $ZZ$ rate required does not exceed the limit of the system spectrum.

\begin{figure}[h]
    \includegraphics[width=0.9\linewidth,keepaspectratio]{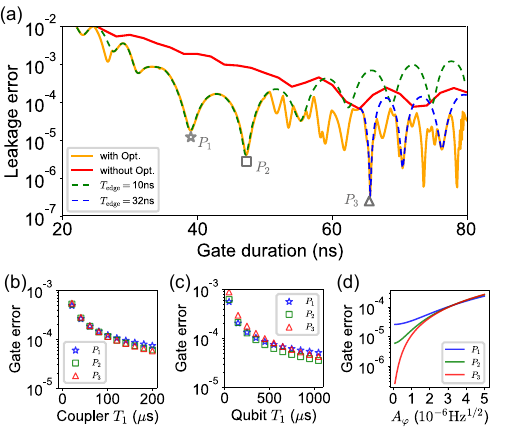}
        \caption{\label{fig:ad2}Optimization of the adiabatic gate pulses and gate errors in the presence of dissipation. (a) Simulation results of the leakage error for various gate pulses. The red line indicates the leakage errors without flat duration optimization, where the flat duration is settled to 0 ns. The orange line presents the minimal leakage errors with the optimization of the flat duration in our simulation. The dashed green and blue line denotes leakage errors with the pulses of 10-ns and 32-ns edge duration, respectively. (b) and (c) represent the CZ gate errors in the presence of coupler's dissipation and qubits' dissipation, respectively. The blue stars, green squares and red triangles respectively denote the pulse 1($P_1$), pulse 2($P_2$), and pulse 3($P_3$), as marked in (a). (d) CZ gate errors in the presence of $1/f$ flux noise with the quasistatic approximation. }
    \end{figure}

Next, we optimize the flat duration to further eliminate the leakage error in the CZ gate. Fig.~\ref{fig:ad2} (a) shows the oscillation of leakage error with a fixed-edge pulse and duration of 10 ns and 32 ns (blue and green curves), and we attribute the oscillations to the interference of the unwanted residual population exchanges in the rising and falling edges. 
Our goal is to optimize the flat duration to induce destructive interference to minimize the leakage population at the end of the pulse. It is notable that for the pulses with a total duration longer than 35 ns, almost all the optimized values $\delta \theta$ are small enough, and no further optimizations are needed to perform the CZ gate, and thus the leakage errors can be directly quantified as CZ gate errors ($\delta \theta=0$). 
In Fig.~\ref{fig:ad2} (a), the orange curve represents the minimum gate errors acquired by sweeping the edge and the flat duration. The CZ gate error 
 can reach the levels below $10^{-4}$, $10^{-5}$, $10^{-6}$ with gate duration less than 40 ns, 50 ns, and 70 ns respectively. 
In comparison,  the red curve illustrates the
gate errors from the unoptimized pulses with a zero flat duration,
which implies that this optimization method is effective, reaching the best error-rate improvement by about three orders
of magnitude.

\subsection{Discussion of the gate performance in practice\label{sec:actual_gate}}
Our adiabatic CZ gate scheme relies on coupler frequency tunability. The implementation only requires that the transmon coupler's zero-flux frequency exceeds the fluxonium $\ket{11}$ computational state energy. This condition provides an operational flexibility, as it eliminates the need for precise parameter targeting.

Our optimization protocol requires accurate characterization of both the ${\mathcal D_{i,j}}$ factor and ZZ interaction strength. Achieving both high speed and fidelity in the gate operation requires strong qubit-coupler coupling, as larger energy gaps enhance adiabatic performance. As a result, it is crucial to design $|J_{c}n_{i,01}n_{c,01}(n_{c,12})|$ as large as possible. For illustration, we include simulation results for a reduced coupling strength ($J_c=400$ MHz) as a comparative benchmark (Fig.~\ref{fig:extral_flat}). Fig.~\ref{fig:ad2}(a) shows the leakage errors of the adiabatic CZ gates with $J_c=600$ MHz, which are generally better than those of the case where $J_c=400$ MHz. Given that $n_{c,01}(n_{c,12})\sim 1(1.4)$ is fixed for transmon, there is still room for adjusting $|J_{c}n_{i,01}|$ to achieve stronger coupling. However, it is notable that both $|J_c|$ and $|n_{i,01}|$ have upper limits in practice. While stronger coupling enables higher gate fidelity, it simultaneously increases unwanted stray couplings (as quantified in Appendix.~\ref{apd:a}), leading to enhanced crosstalk. Besides, increasing $|n_{i,01}|$ degrades the qubit protection from the noisy environment, and thus has negative impact on qubit coherence time. In conclusion, a trade-off between gate fidelity and scalability must be considered to achieve a high-performance large-scale quantum processor.
 
The constant-leakage-rate edge in our simulations is based on the priori knowledge of the system Hamiltonian. However, parameter variations are almost inevitable, considering the fabrication mistargeting and complicated multi-couplings in a large-scale quantum device. In addition, due to the limited bandwidth of flux lines, waveform distortion is also unavoidable. Therefore in practice these two effects will lead to deviations in the final pulse from the actual constant-leakage-rate trajectory. Further work may include treatment of calibrating the edge shape in an experimental manner, which is outside the scope of this work. 

Another crucial gate error source is derived from the decoherence process. In the presence of dissipation, suppressing the charge-noise sensitivity of transmon is essential since the transmon here works at a low frequency range of $\sim$3 GHz during the gate. Therefore in our simulation a small charge energy $E_C=180$ MHz is chosen to improve the ratio $E_J/E_C$. We choose three optimized pulses with gate duration (infidelity) 39.5 ns ($2.5\times 10^{-5}$), 47 ns ($5.8\times 10^{-6}$), 65.5 ns ($1.5\times 10^{-7}$) in our estimation of the incoherent errors, marked in Fig.~\ref{fig:ad2}(a). The edge duration of pulses $P_1$ and $P_2$ is 10 ns, while that of $P_3$ is 32 ns. We model the dissipation process with master equations (see Appendix.~\ref{ap:ies} for details) and simulate the CZ gate fidelity versus coupler's and qubits' $T_1$. From Fig.~\ref{fig:ad2}(b) and (c), we observe a trend that gate error is dominated by coherent error when $T_1$ is long enough, while for short $T_1$, the error from dissipation dominates and the fidelity is lower with a longer gate duration. To achieve a CZ gate fidelity above 99.99\%, the lower bounds of the qubits' $T_1$ and the coupler's $T_1$ are about 500 $\mu s$, and 150 $\mu s$, corresponding to a dielectric quality factor $Q_\mathrm{diel}$ of $1.1\times10^7$ and $3.5\times10^6$, respectively.

Another main incoherent error originates from the dephasing process. We assume the dephasing is dominated by the flux noise from the coupler flux line and neglect the fluxoniums' flux noise. This is reasonable as the fluxonium coherence times are protected to the first order in flux noise at integer flux quanta. The noise power spectrum
of flux noise generally exhibits a “quasiuniversal” dependence in practice, $S_{\varphi}(f) = A_{\varphi}^2/f^\alpha$. In this work we simplify the $1/f$ noise model by using a quasistatic noise assumption to keep only the lowest frequency component of $1/f$ noise. That is, assuming the noise as a stochastic but static flux bias $\delta\varphi_{\rm ext,c}$, which follows a Gaussian distribution with a standard deviation of the same value for a $1/f$ noise ($S_{\varphi}(f) = A_{\varphi}^2/f$). In this way we find that the quasi-static flux-noise-induced error can reach below $\times 10^{-4}$, given a typical $1/f$ noise intensity $A_{\varphi}=1\times 10^{-6}{\rm Hz}^{-\frac{1}{2}}$ (Fig.~\ref{fig:ad2}(d)).

\section{MICROWAVE-ACTIVATED CZ GATE\label{sec:microwaveCZ}}
A microwave-activated CZ gate typically requires lifted degeneracy by qubit couplings. As a partially-protected qubit, fluxonium's charge matrix element for $0-1$ transition is constrained to suppress the susceptibility to electric noises, which also limits the capacitive coupling strength in the FTF system. To implement a fast two-qubit gate, an alternative solution is to take advantage of the large matrix elements of the transitions involving non-computational states. As for integer fluxoniums, the plasmon transition $0-3$ or $1-4$ is on-demand to split the coupler energy levels  due to its charge matrix element of
$\sim 0.5$ and transmon-like frequency. 
\subsection{Gate principle}

In this section we showcase a microwave-activated CZ gate scheme by selectively driving the transition between the ground state and the coupler's first excited state in a full Rabi cycle to gain a geometric $\pi$ phase. This gate scheme is based on the energy splitting of the coupler's first excited state, which is repelled by the third excited states of both fluxoniums. In our calculation, we change the coupler's effective $E_J$ by tuning external flux following Eq.~\ref{eq:ftf1}. In practice, one should avoid large flux tunability to suppress flux noise. For example, it is better to use asymmetric-junction transmons as they are less sensitive to flux or simply use fixed-frequency transmons. The effective Hamiltonian can be described as 
\begin{equation}
    H_{\mathrm{eff}}'/h = \chi^c_{00}\ket{\widetilde{010}}+\chi^c_{01}\ket{\widetilde{011}}+\chi^c_{10}\ket{\widetilde{110}}+\chi^c_{11}\ket{\widetilde{111}},
\end{equation}
where $\chi^c_{ij}$ is the coupler frequency shift at the fluxonium qubit state $\ket{ij}$, and $\chi_{00}$ is highly detuned from $\chi_{01}$, $\chi_{10}$, and $\chi_{11}$. Similar gate scheme has been demonstrated experimentally in Ref.~\cite{ld}, where the conventional fluxoniums are driven by the two corresponding charge lines simultaneously. 
\begin{figure} 
    \includegraphics[width=0.9\linewidth,keepaspectratio]{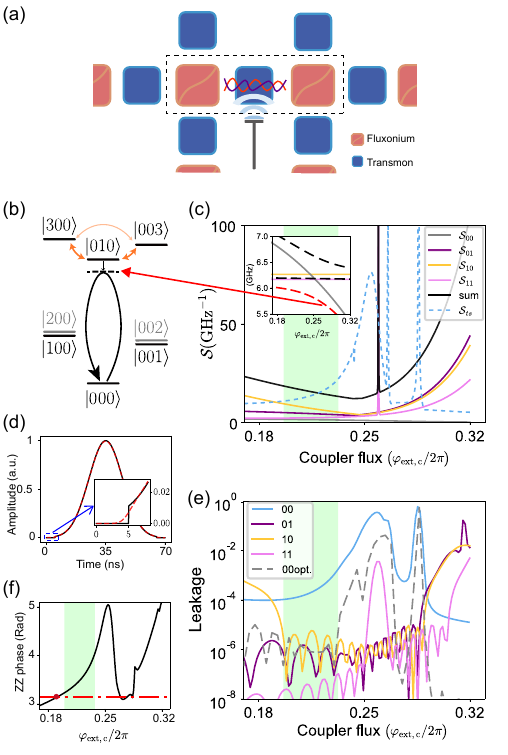}
    \caption{\label{fig:mwfig}Microwave-activated CZ gate scheme simulations. The Hamiltonian parameters in simulation are: $E_{C,1(2)}=1.2$ GHz, $E_{J,1(2)}=6.1(6)$ GHz, $E_{L,1(2)}=0.17(0.15)$ GHz, $E_{C,c}=300$ MHz, $E_{J,c}=25$ GHz, $J_{C,1(2)}=450$ MHz, and $J_{12}=120$ MHz. (a) Sketch of the coupler-assisted microwave-activated CZ gate architecture. The microwave pulse is delivered via the couplers' dedicated charge drive line to the FTF system. (b) Schematic of the microwave-activated CZ gate scheme. The significant selectivity in driving the target transition arises from the simultaneous repulsion of the third excitation states of the fluxoniums. Here the target state is identified as the lowest dressed state within the three-level interaction subspace, as depicted by the red dashed line in the inset of subfigure (c). (c) Calculated $\mathcal{S}$ parameters VS coupler flux bias $\varphi_{\rm ext,c}$. The black line represents the sum of all $\mathcal{S}_{ij}$ values. The light green region denotes the parameter space where unwanted transitions are effectively suppressed. The inset illustrates the frequencies of the corresponding states. The dashed lines represent the three eigenenergies of the dressed states in the three-level interaction subspace, with the red line indicating the target state. The orange, pink, and gray lines represent the frequencies of the bare states $\ket{300}$, $\ket{010}$, and $\ket{003}$, respectively. (d) Diagram of the envelop of the microwave pulse with 60-ns gate duration. The standard deviation for the Gaussian filter kernel is 1 ns, and the total width is 4 ns. (e) Population leakages of the computational states VS coupler flux, after the on-resonance $2\pi$ rotation pulse. The solid lines indicate the leakages with the pulse amplitude of $1/(\mathcal{A}N_{T})$, while the dashed grey line represents the suppressed leakage of $\ket{00}$ by optimizing the pulse amplitude. (f) Extracted $ZZ$ phases VS coupler flux after applying on-resonant $2\pi$ rotation pulse.} 
\end{figure} 
The microwave-activated gate in our scheme utilizes driving through the coupler's dedicated charge line (Fig.~\ref{fig:mwfig}(a)), offering distinct advantages over alternative approaches. Conventional methods relying on simultaneous drives to computational qubits typically require complex calibration protocols with cancellation pulses, introducing significant overhead for gate optimization. In contrast, our single-line control architecture inherently suppresses crosstalk while substantially simplifying the calibration process. Direct qubit driving presents a significant challenge as it preferentially excites high-energy fluxonium transitions (such as the 0-3 transition) over the target coupler transition. Our dedicated coupler line architecture effectively suppresses these leakage channels while maintaining gate fidelity. This advantage facilitates the practical implementation of integer-fluxonium qubits in scalable quantum processors.

The main idea of this gate is to selectively drive the target transition while minimizing the influence on other transitions. To quantify the impact of the unwanted transitions, we define a parameter \(\mathcal{S}_{ij}\) as follows:
\begin{equation}
    \mathcal{S}_{ij} = \max_m \left\{\frac{N_{ij,m}}{N_{T} \left( |E_{ij,m} - E_T| \right)}\right\},
\end{equation}
where \(i,j,m\) represent the transition between \(\ket{\widetilde{i0j}}\) ($i,j=0,1$) and \(\ket{\widetilde{m}}\), with \(m\) denoting the \(m\)-th eigenstate of the FTF system. The index \(T\) corresponds to the target transition, \(\ket{\widetilde{000}} \to \ket{\widetilde{010}}\). \(N_{ij,m} = |\langle \widetilde{i0j} | \hat{n}_c | \widetilde{m} \rangle|\) represents the charge drive matrix element for the transition between \(\ket{\widetilde{i0j}}\) and \(\ket{\widetilde{m}}\). \(E_{ij,m} = |E_{ij} - E_{m}|\) denotes the energy gap between the states \(\ket{\widetilde{i0j}}\) and \(\ket{\widetilde{m}}\). \(E_T\) and \(N_T\) represent the energy gap and the charge drive matrix element associated with the target transition, respectively. When calculating \(\mathcal{S}_{00}\), the state \(\ket{\widetilde{010}}\) should be excluded from the set of \(m\) states. 
The parameter \(\mathcal{S}_{ij}\) quantifies the relative driving strength of unwanted transitions compared to the target transition. A large value of \(\mathcal{S}_{ij}\) indicates that the unwanted transitions are more easily provoked, which may produce larger leakage errors and ac-Stark shifts. Similarly, we can also define the parameter $\mathcal{S}_{ts}$ as 
\begin{equation}
    \mathcal{S}_{ts}=\max_m \left\{\frac{N_{ts,m}}{N_{T} \left( |E_{ts,m} - E_T| \right)}\right\},
\end{equation}
where the index $ts$ denotes the target state $\ket{\widetilde{010}}$. This parameter quantifies the relative strength of unwanted transitions originating from the target state. To evaluate the gate scheme and demonstrate the effectiveness of the $\mathcal{S}$ parameters, we apply gate pulses at different coupler flux. Each pulse features a Gaussian envelope with a duration of 60 ns and a standard deviation of $\sigma=15$ ns. Our simulation also includes a 5 ns idle period at both the start and the end of the pulse. The pulse amplitude is set to $1/(\mathcal{A}N_{T})$, where $\mathcal{A}$ is the area of the pulse envelope, to achieve a $2\pi$ rotation. The driving frequency is the target transition frequency. The initial states of the simulation are the four computational states $\ket{\widetilde{000}}$, $\ket{\widetilde{001}}$, $\ket{\widetilde{100}}$ and $\ket{\widetilde{101}}$. As shown in Fig.~\ref{fig:mwfig} (e), the final leakage population of the four initial states exhibits qualitative agreement with the corresponding $\mathcal{S}$ parameters. Furthermore, the leakage is suppressed effectively, with the primary leakage attributed to the leakage to the target state. Notably, the leakage from the $\ket{00}$ state is mainly caused by the ac-Stark shift of the target state, resulting the deviation of the $2\pi$ rotation. However this leakage can be easily suppressed by optimizing the amplitude of the gate pulse (dashed grey line). Thus we can identify an optimal coupler frequency. As shown in Fig.\ref{fig:mwfig} (e), the overall leakage can be suppressed in the light-green region, corresponding to a coupler frequency range of approximately 300 MHz, which makes it possible to substitute the frequency-tunable transmon with a fixed-frequency transmon in practice. Within this range, the coupler's first excited state is highly hybridized with the fluxoniums's third excited states, leading to a significant detuning of $\chi_{00}$ from $\chi_{01}$, $\chi_{10}$, and $\chi_{11}$. Additionally, there are neither near-resonant unwanted transitions (sharp peaks) nor severe off-resonant excitations (broad peaks). 

\subsection{Optimization for ac-Stark shift phases}
Given the population leakages are suppressed to the level of $10^{-6}$, next we focus on the $ZZ$ phase. We extract the $ZZ$ phase from the final states in the previous simulations. The result indicates a significant deviation from the ideal value of $\pi$, as shown in Fig.~\ref{fig:mwfig} (f). The phase deviation can be described as
\begin{equation}
    \delta\theta_{{\rm zz}}=\theta_{11}-\theta_{10}-\theta_{01}+\theta_{00}-\pi,
\end{equation}
where the $\theta_{ij}$ ($i,j=0,1$) is the phase obtained in respect of the state $\ket{ij}$. 
This deviation can be attributed to the inevitable ac-Stark shifts of the computational states, which arise from the off-resonant drives of the unwanted transitions. Given that the ac-Stark shifts over long time scales are inversely proportional to the gate duration for the pulses with a fixed $\sigma/T_g$ ratio, using longer-duration pulses can mitigate these shifts, though this approach comes with significant drawbacks. An alternative method to reduce the phase deviation is to adjust the coupler flux to achieve destructive interference of the ac-Stark shifts (indicated by the red dot in Fig.~\ref{fig:mwfig} (f)), which is effective with a frequency-tunable coupler. In practice, however, this method is ineffective for fixed-frequency couplers due to fabrication uncertainties. Additionally, this method does not guarantee that the leakage and the phase errors can be minimized at the same flux. Therefore, an on-site solution is more favorable. 

Here we propose to adjust the $ZZ$ phase by introducing a detuning of the driving frequency. This can be understood by a simplified spin model with an off-resonant drive, where only the target transition is considered. In the frame of the drive, the Hamiltonian takes the form of
\begin{equation}
    H_s/h = \frac{\Delta}{2} \sigma_z+\frac{\Omega}{2}\sigma_x,
\end{equation}
where $\Delta=f_t-f_d$ is the detuning between the target transition frequency $f_t$ and the driving frequency $f_d$. $\Omega$ is the driving strength. In this model, any $2\pi$ rotation will result in a phase of $\pi$ for the initial state $\ket{0}$. When the system is transformed back to the lab frame, an additional phase shift of $\pi\Delta T_g$ is obtained. The total phase is therefore a geometric phase that equals half the solid angle of the spin trajectory on the Bloch sphere. For a square pulse, the phase tunability is $\delta\theta_{00}/\Delta=\pi T_g$. For a Gaussian pulse as depicted in (Fig.~\ref{fig:mwfig}(d)), the phase tunability is approximately $0.663\pi T_g$, which means a maximal $\pi$ phase deviation can be compensated with a detuning of approximately 25 MHz($T_g = 60~\mathrm{ns}$). Considering the full Hamiltonian, the modification on the drive frequency also sightly changes the ac-Stark shifts, but one can always fine-tune the drive frequency to completely compensate the $ZZ$ phase deviation. 

\subsection{Gate error after optimization}
\begin{figure} 
    \includegraphics[width=0.9\linewidth,keepaspectratio]{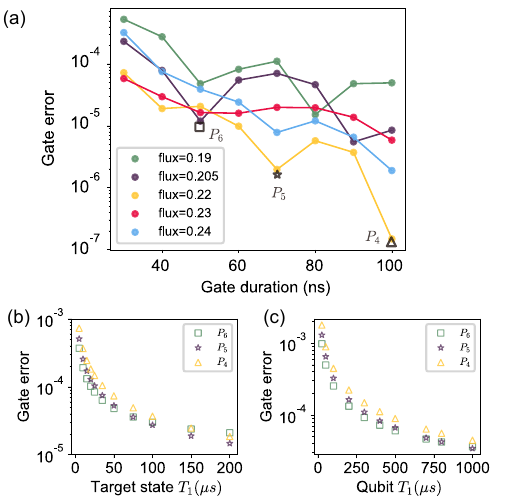}
    \caption{\label{fig:mw2}Gate fidelities after ac-Stark phases cancellation. (a) Gate errors with various coupler fluxes and durations, where each gate pulse is obtained with optimizations of both driving frequency and amplitude. The 'square', 'star', and 'triangle' mark the gate pulse $P_6$, $P_5$ and $P_4$ in the dissipation simulation in (b) and (c). (b) gate error VS target state's $T_1$. (c) gate error VS qubits' $T_1$, where in our simulation we assume the same values of $T_1$ for both fluxoniums.} 
\end{figure} 
 We analyze gate errors across various flux values and durations. The results in Fig.~\ref{fig:mw2}(a) indicate that the gate fidelity is strongly dependent on the coupler frequency. By finely tuning the coupler flux, we can achieve a gate error at \(1 \times 10^{-5}\) level with a gate duration of 50 ns, and even a lower level of  \(1 \times 10^{-6}\)  and \(1 \times 10^{-7}\) with a duration of 70 and 100 ns, respectively. For a fixed-frequency coupler, it is feasible to realize a 50 ns CZ gate with a small coherent error on the order of \(10^{-5}\).

To investigate gate fidelity in the presence of dissipation, we first consider the impact of the target state's \(T_1\), which arises from the dissipative processes for both the coupler and the fluxoniums. We model this effect as the decay of the target state \(\ket{\widetilde{010}}\) to the ground state. We simulate the corresponding gate errors for three durations: 50 ns, 70 ns, and 100 ns, as marked in Fig.~\ref{fig:mw2}(a). The results are presented in Fig.~\ref{fig:mw2}(b). Additionally, we simulate gate fidelities with varying qubits' \(T_1\), illustrated in Fig.~\ref{fig:mw2}(c). These findings suggest that to implement a CZ gate with an error below \(1 \times 10^{-4}\), the lower bound for the target state \(T_1\) is approximately 25 $\mu$s; while for the fluxoniums, the lower bound is around 500 $\mu$s, corresponding to a $Q_{\rm diel}$ of about $1\times10^5$.

\subsection{Comparision with the adiabatic CZ gate}
Here we include a comparison of the two gate schemes to better highlight their respective advantages and limitations.

\paragraph{Parameter targeting} The adiabatic CZ gate offers significant operational flexibility, as it imposes no stringent requirements on circuit parameters. This design leverages the tunable coupler to enable dynamic control of $ZZ$ interactions, allowing for their programmable suppression or enhancement as needed. In contrast, the microwave-activated CZ gate scheme places more stringent requirements on the design of fluxonium $\ket{3}$ states. The design presents a trade-off: stronger coupling enables faster gate operations but simultaneously imposes stricter precision requirements for gate transition frequencies. When significant detuning exists between the fluxonium $\ket{3}$ states and the transmon, achieving effective gate operation requires careful engineering of the qubit-coupler coupling to properly split the gate transition frequency.
\paragraph{Decoherence} Flux noise poses a significant challenge for adiabatic gates, as the transmon must be biased at a flux-sensitive operating point while relying on fast-flux lines to generate the necessary flux pulses. In contrast, the microwave-activated scheme does not rely on flux tunability in principle. This approach enables the use of fixed-frequency transmons or flux-insensitive transmons along with heavily-filtered flux lines. One potential concern involves the coherence properties of the $\ket{3}$ state, which remain relatively unexplored experimentally. It is currently unclear whether the $\ket{3}$ state can achieve coherence times comparable to those of the transmon's computational states.
\paragraph{Hardware overhead} Both schemes require a dedicated control line for coupler operation, resulting in similar hardware overhead in their basic implementations. However, preserving coupler tunability in the microwave-activated scheme necessitates an additional flux line.
\paragraph{Spectator qubit effect} 
In large-scale circuits, spectator qubits can introduce additional gate errors. Here we focus on impacts of next-nearest qubits in the two gate schemes. For adiabatic CZ gates, spectator states create anti-crossings in the adiabatic trajectory with MHz-scale gaps. This requires pulse-shape optimization to minimize leakage errors during flux sweeps through these anti-crossings. In the case of microwave-activated gates, spectator coupling splits the coupler transition frequency, leading to both leakage and phase errors. Our analysis shows this effect remains manageable - for instance, a 10 MHz stray coupling induces a modest 0.15 MHz shift in the coupler frequency, enabling the microwave-activated gates to maintain fidelities exceeding $99.99\%$ (see Appendix~\ref{appendix:map_gate} for detailed analysis).

\section{IF-T-F SYSTEM\label{sec:IFTF}}
In this section, we discuss the IF-T-F system, where an integer fluxonium qubit and a conventional fluxonium qubit are coupled via a transmon coupler. Due to the significant difference in operating frequencies between the two types of fluxoniums, this architecture effectively mitigates single-qubit gate crosstalk between neighboring qubits and alleviates the frequency crowding issues.

\begin{figure} 
    \includegraphics[width=0.9\linewidth,keepaspectratio]{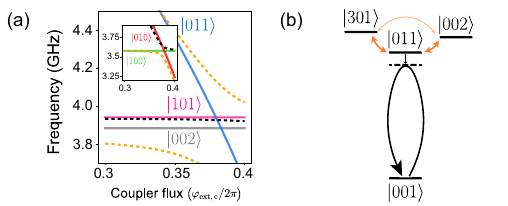} 
    \caption{\label{fig:IFHIF} Revised CZ gate schemes in the IF-T-F system. (a) Frequencies of bare states (solid lines) and the eigenstates (dashed black and orange lines) in the double-excitation manifold VS coupler flux $\varphi_{\rm ext,c}$ for the revised adiabatic CZ gate scheme. The non-trivial $ZZ$ phase is achieved by shielding the repulsion of the state $\ket{011}$ with the state $\ket{002}$. The Hamiltonian parameters are: $E_{C,1(2)}=1.1$ GHz, $E_{J,1(2)}=4 (3.6)$ GHz, $E_{L,1(2)}=0.2 (0.5)$ GHz, $E_{C,c}=250$ MHz, $E_{J,c}=20$ GHz, $J_{C,1(2)}=500$ MHz, and $J_{12}=200$ MHz. Inset shows the frequencies of those states in the one-excitation manifold.(b) Schematic revised microwave-activated CZ gate scheme, where the integer fluxonium's $0-3$ transition is replaced with the conventional fluxonium's $1-2$ transition. The $ZZ$ phase is acquired by the computational state $\ket{01}$.}
\end{figure}

Here we introduce two-qubit gate schemes for this system. Firstly, the adiabatic CZ gate scheme described above can not directly apply to the IF-T-F system. This is because the frequency of the conventional fluxonium is typically hundreds of megahertz, which leads to a very low frequency of the computational state $\ket{11}$. Consequently, it is impossible to achieve simultaneous near-resonance conditions for $\ket{010}-\ket{100}$ and $\ket{020}-\ket{101}$ (the state label follows the order: integer fluxonium, transmon and conventional fluxonium). However, one can use fluxonium's second excitation state to achieve the non-trivial energy level repulsion. One approach is to set the frequency of $\ket{002}$ close to that of $\ket{101}$ to frozen the frequency shift of $\ket{101}$. In other words, $\ket{101}$ is shielded by $\ket{002}$ from the repulsion of $\ket{011}$, since the interaction strength between $\ket{011}$ and $\ket{002}$ ($\sim  n_{\mathrm{cf},12}$) is larger than that between $\ket{011}$ and $\ket{101}$ ($\sim n_{\mathrm{if},01}$), $\ket{011}$ will be pushed away when it approaches $\ket{101}$ as shown in Fig.~\ref{fig:IFHIF}(a). In this way, the adiabatic CZ gate can still be optimized by the two-step method (presented in Sec.~\ref{optad}). We note that this gate scheme is still feasible for heavy conventional fluxoniums, which are characterized by a very small $n_{01}$ and are expected to exhibit enhanced coherence times. In contrast, heavy integer fluxoniums are not suitable to this gate scheme due to the requirement for a strong coupling between $\ket{100}$ and $\ket{010}$. 

The microwave-activated CZ gate scheme, however, can still work for the IF-T-F system by setting the conventional fluxonium transition frequency of $\ket{1}$ and $\ket{2}$ near resonant with the integer fluxonium transition frequency of $\ket{0}$ and $\ket{3}$. In this scenario the target transition should be $\ket{\widetilde{001}}$ to $\ket{\widetilde{011}}$, and the geometric $\pi$ phase is acquired via $\ket{\widetilde{001}}$ as indicated in Fig.~\ref{fig:IFHIF}(b). Besides, this gate scheme applies to both heavy integer fluxoniums and heavy conventional fluxoniums.

\section{DISCUSSION\label{sec:dis}}
As mentioned earlier, the leakage error during the gate can be properly handled by designing the circuit parameters and fine-tuning the control pulse. In practice, the gate error is usually more constrained by the incoherent error. Our simulation in Sec.~\ref{sec:actual_gate} includes the effects of the $T_1$ of the qubits and the coupler and the $1/f$ flux noise. However, the noise spectrum of flux noise can have a different frequency dependence in the experiment, and affect the gate performance in a more detrimental way, which is also observed in a previous study \cite{PhysRevApplied.22.024057}. To get a rough estimation on the gate error, one can measure the coupler's $T_{\phi}$ under the influence of flux noise experimentally and assume the gate error has the same dependency \cite{PhysRevApplied.22.024057} on $T_1$ and $T_\phi$. Notably, the coupler's $T_1$ imposes a smaller impact on the gate performance in comparison to the qubit's $T_1$. In addition to the decoherence process, thermal excitation of the coupler also contributes to the gate error. The coupler frequency is near 6 GHz for both gate schemes, indicating the thermal population of excited coupler states remains below $1\times10^{-4}$ at 30 mK. This imposes a practical limit on CZ gate error rate of $1\times10^{-4}$.

Integer fluxoniums operating at zero flux have distinct spectrum structure compared with conventional fluxoniums operating at half flux quanta. The qubit frequency is several gigahertz at zero flux, which is close to the transmon qubit. This feature renders the adiabatic flux gate similar to that in a full-transmon architecture in certain aspects. On the other hand, the microwave-activated CZ gate takes advantage of the high energy transitions, where the large anharmonicity of the integer fluxoniums distinguishes the system from that of transmon qubits. In both cases, our proposal makes for  both high-fidelity and scalable gate schemes. First of all, the static $ZZ$ interaction can be suppressed and eliminated at proper flux bias. The far-detuned $\ket{2}$ states of the integer fluxonium alleviate concerns about the small-gap anti-crossings in the adiabatic trajectory. The total error rate for the flux gate can reach the level of $10^{-4}$ in consideration of leakage and decoherence. For the microwave-activated gate, our gate scheme supports single-line control, which suppresses crosstalk and simplifies calibration. These advantages enable the practical application of integer fluxonium qubits in large-scale quantum circuits. Besides the advantages of the extended frequency range and relaxed requirements on flux tunability, integer fluxonium is also a promising qutrit \cite{PhysRevLett.113.230501} candidate due to its protected high energy state. Therefore it can be anticipated that scaling up the integer fluxonium system will bring numerous new opportunities to this field. 

To conclude, we propose the high-fidelity scalable coupling schemes for integer fluxoniums with a transmon coupler and analyze the gate performance. The coupling scheme can efficiently suppress static $ZZ$ interaction. Furthermore, with the state-of-the-art median coherence time $T_1$ of transmons surpassing 100 microseconds in a two-dimensional configuration\cite{Xiang2024}, and the anticipated enhancement in the dielectric quality factor of integer fluxonium, we are optimistic that it is feasible to achieve a CZ gate fidelity exceeding 99.99\% in experimental settings utilizing our proposed gate schemes. Our proposal offers valuable insights to construct large scale circuits composed of integer fluxoniums in the future.
\section*{Acknowledgments}

This work is supported by Innovation Program for Quantum Science and Technology (Grant No.2021ZD0301704), China Postdoctoral Science Foundation (Grant No.2023M742005) and Shuimu Tsinghua Scholar Program. 

\appendix
\section{Full capacitance analysis and system Hamiltonian\label{apd:a}}
Here we introduce various FTF circuits, derive the Hamiltonian parameters and discuss their relationships. 
\subsection{1D chain}

\begin{figure*}[t!]
\includegraphics[width=0.9\linewidth,height=5cm,keepaspectratio]{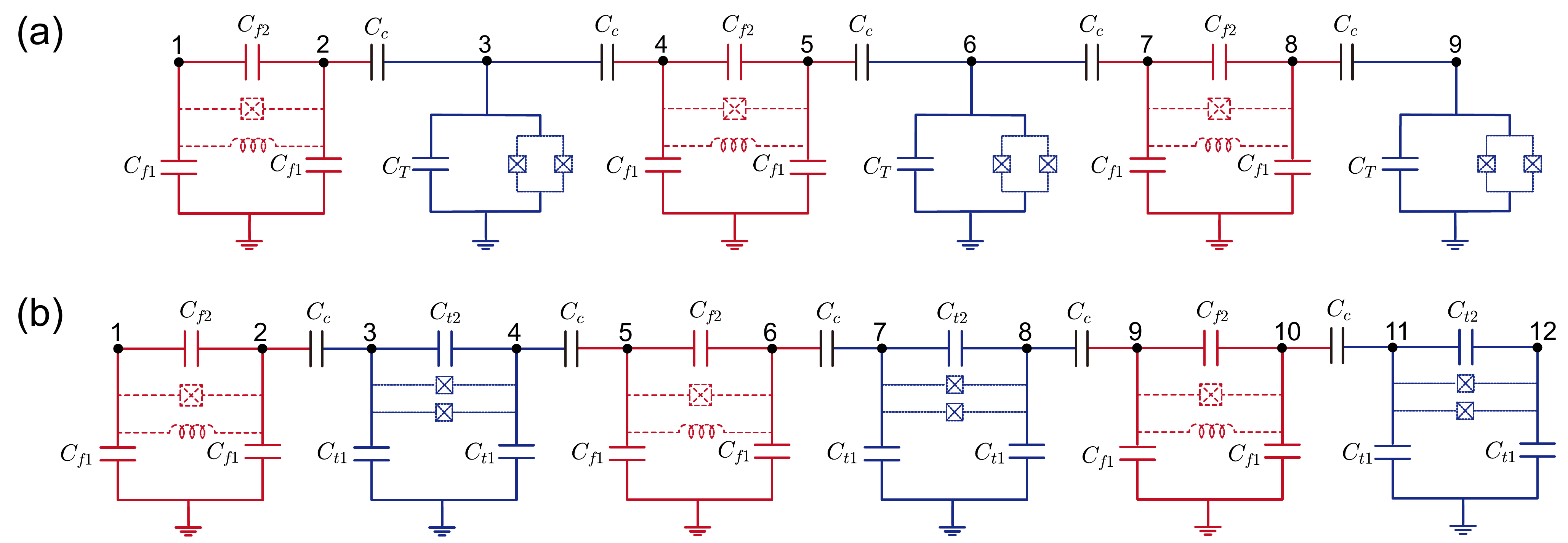}
\caption{\label{fig:circuit1}Circuit models of FTF circuits of one-dimensional scaling archtecture. The capacitance network
is simplified for the purpose of a theoretical analysis,
with no direct fluxonium-fluxonium coupling capacitance. (a) FTF circuit with grounded transmons.  (b) FTF circuit with differential transmons.}
\end{figure*}
Starting from the node methods, we first consider a one-dimension fluxonium qubit chain with grounded transmons as couplers. The circuit diagram is shown in Fig.~\ref{fig:circuit1}, and here we neglect the small and qualitatively unimportant capacitance in this analysis. The capacitance matrix of this circuit can be written as 
\begin{widetext}
    \begin{equation}
    C = \begin{pmatrix}
          C_f    & -C_{f2}  &0 & 0& 0&0 &0 &0 \\ 
         -C_{f2} &  C_f+C_c &-C_c &0 &0 &0 & 0&0 \\
                0 & -C_c & C_T+2C_c &-C_c &0 & 0& 0& 0\\
                0 & 0  & -C_c  &C_f+C_c &-C_{f2} & 0& 0&0 \\
                 0 & 0 & 0&-C_{f2} & C_f+C_c&-C_c  & 0&0 \\
                 0   & 0  &0 &0 &-C_c & C_T+2C_c&-C_c &0\\
                 0   &0   &0 &0 &0 & -C_c & C_f+C_c &-C_{f2}\\
                  0  &0   &0 &0 &0 &0 &-C_{f2} &C_f\\
    \end{pmatrix},
    \end{equation}
\end{widetext}
where we define $C_f=C_{f1}+C_{f2}$ for convenience. To extract the differential mode used as the qubit modes, a variable transformation is needed to modify the capacitance matrix as $\tilde{C}=(M^T)^{-1}CM^{-1}$ with 
\begin{widetext}
\begin{equation}
    M = \begin{pmatrix}
          1    & 1  &0 & 0& 0&0 &0 &0 \\ 
         1 &  -1 &0 &0 &0 &0 & 0&0 \\
                0 & 0 & 1&0 &0 & 0& 0& 0\\
                0 & 0  & 0 &1 &1 & 0& 0&0 \\
                 0 & 0 & 0&1 & -1&0  & 0&0 \\
                 0   & 0  &0 &0 &0& 1&0 &0\\
                 0   &0   &0 &0 &0 & 0& 1&1\\
                  0  &0   &0 &0 &0 &0 &1 &-1\\
    \end{pmatrix}.
    \end{equation}
    \end{widetext}
To get rid of the complex exact form of the elements in $ \tilde{C}^{-1}$, a further simplification can be applied based on the relation $C_{f1}\approx C_{f2}\approx C_c \ll C_T$. Thus, 
\begin{widetext}
\begin{equation}\label{eqa3}
    \begin{aligned}
    \tilde{C}^{-1}[2,2]&\approx \frac{C_c+2C_{f1}}{C_c C_f+C_f^2-C_{f2}^2} \approx \frac{1}{\frac{5}{3}C_s},\\
    \tilde{C}^{-1}[5,5]&\approx \frac{2}{C_c+C_f+C_{f2}} \approx \frac{1}{2C_s},\\
    \tilde{C}^{-1}[2,3]&\approx \frac{-C_c C_{f1}}{C_T(C_c C_f+C_f^2-C_{f2}^2)}\approx -\frac{1}{5}\frac{1}{C_T}\frac{C_c}{C_s},\\
    \tilde{C}^{-1}[5,3]&\approx \frac{C_c}{C_T(C_c+C_f+C_{f2})}\approx \frac{1}{4}\frac{1}{C_T}\frac{C_c}{C_s},\\
    \tilde{C}^{-1}[2,5]&\approx \frac{C_c^2C_{f1}^2}{C_T(C_c C_f+C_f^2-C_{f2}^2)^2}\approx \frac{1}{25}\frac{1}{C_T}\frac{C_c}{C_s}\frac{C_c}{C_s},\\
    \tilde{C}^{-1}[3,6]&\approx \frac{C_c^2C_{f2}^2}{C_T^2(C_c^2+2C_c C_f +C_f^2-C_{f2}^2)}\approx \frac{1}{8}\frac{C_s}{C_T^2},\\
    \tilde{C}^{-1}[3,9]&\approx \frac{C_c^4C_{f2}^2}{C_T^3(C_c^2+2C_c C_f +C_f^2-C_{f2}^2)^2}\approx \frac{1}{64}\frac{C_s^2}{C_T^3},\\
    \tilde{C}^{-1}[2,6]&/\tilde{C}^{-1}[2,3]\approx \frac{C_c^2 C_{f1}}{C_T(C_c^2+2C_c C_f+C_f^2-C_{f2}^2)}\approx \frac{1}{8}\frac{C_s}{C_T},\\
    \tilde{C}^{-1}[2,8]&/\tilde{C}^{-1}[2,6] = \frac{C_c C_{f1}}{C_c C_f+C_f^2-C_{f2}^2}\approx \frac{1}{5};
    \end{aligned}
\end{equation}
\end{widetext}
where $C_s$ represents the small capacitance of the same order of magnitude with $ C_{f1}$ , $C_{f2} $ and $ C_c $; the indices $2,5,8$ represent the three adjacent qubit modes while $3,6$ represent the two grounded transmon coupler modes. Ignoring the sum modes that are assumed to have a minimal effect on the dynamics, the Hamiltonian of the system is:
\begin{equation}\label{eqa4}
    \begin{aligned}
    H/h = &\sum_{i=1,2,3}[4E_{C,i}\hat{n}_i^2+\frac{1}{2}E_{L,i}\hat{\phi}_i^2
    -E_{J,i} \cos(\hat{\phi}_i-\phi_{\rm ext,i})]\\
    + &\sum_{j=1,2}[4E_{C,cj}\hat{n}_i^2-E_{J,cj} \cos(\hat{\phi}_j-\phi_{\rm ext,cj})]\\
    +&J_{12}\hat{n}_1\hat{n}_2+J_{1c}\hat{n}_1\hat{n}_{c1}++J_{2c}\hat{n}_2\hat{n}_{c1}\\
    +&J_{13}\hat{n}_1\hat{n}_3+J_{1c2}\hat{n}_1\hat{n}_{c2}+J_{cc}\hat{n}_{c1}\hat{n}_{c2}+J^\prime_{c1c3}\hat{n}_{c1}\hat{n}_{c3}\\ +& \cdots.
    \end{aligned}
    \end{equation}
The terms in the first line are the Hamiltonians of three fluxonium qubits (qubit 3 is the spectator qubit); those in the second line are the Hamiltonians of the two transmon couplers, while in third line are the target coupling terms, and in the last line are the unwanted stray coupling terms, where we neglect the unimportant symmetrical coupling terms for demonstration purpose. 

The parameters in the system Hamiltonian of Eq.~\ref{eqa4} can be derived by the elements in the $\tilde{C}^{-1}$, where the diagonal terms give charging energies and the non-diagonal terms give coupling strengths. The last two terms in Eq.~\ref{eqa3} imply  a suppression of the unwanted next-neighbor couplings by two orders of magnitude (assuming $C_T\approx 10C_s$) compared to the target adjacency couplings, which is in a similar order in a transmon qubit-transmon coupler architecture. The third and forth terms implies an approximately constant ratio of $5$ between $J_{1c}$($J_{2c}$) and $J_{12}$.
From the point of designing the circuit capacitance parameters in a scalable manner, the upper bound of $C_c$ is limited by the fluxonium capacitive loading issue and the unwanted couplings' strength. Note that the effective fluxonium's charging energy $E_{C,f}=e^2\tilde{C}^{-1}[n,n]/2$ with $n=2,5,8$; and the fluxonium self-capacitance $C_F=\frac{1}{2}(C_{f}+C_{f2})=\frac{3}{2}C_s$, which corresponding to $1/\tilde{C}^{-1}[2,2]$ when $C_c=0$. The non-trivial coupling capacitance $C_c$ increases the effective fluxonium's capacitance and thus reduces the charging energy, which may breaks the parameters' constraints for building a fluxonium of good performance. Besides, it is obvious that the unwanted couplings (e.g., $J_{13}\propto\tilde{C}^{-1}[2,8],J_{1C2}\propto\tilde{C}^{-1}[2,6],J_{cc}\propto\tilde{C}^{-1}[3,6]$) and quantum crosstalk also increase with larger $C_c$. On the other hand, a fast and high-fidelity two-qubit gate generally demands a large coupling strength between qubits and couplers (a large $C_c$), thus bring about a trade-off between two-qubit gate performance and scalability.  

Replacing the grounded transmon couplers with differential transmon couplers as shown in Fig.\ref{fig:circuit1} (b), the capacitance matrix $\tilde{C}$ can be obtained and investigated using the similar method as outlined above for the grounded transmons case. There are three main differences between the differential transmon coupler case and the grounded transmon coupler case. First, the effective coupling strength is reduced by approximate $1/2$ compared to the grounded transmon case. Second, with an additional designable parameter for the coupler (the ratio between inter-antenna capacitance and grounding parasitic capacitance of the differential transmon, i.e., $C_{t2}/C_{t1}$), the ratio of the qubit-coupler coupling strength to the qubit-qubit coupling strength, as well as the ratio between next-nearest-neighbor and nearest-neighbor coupling strengths, is effectively tunable. Third, in the gauge setting $J_{1c}> 0,J_{2c}> 0$, the differential transmon case gives $J_{12}< 0$ while the grounded transmon case gives $J_{12}> 0$.

\begin{table}[h]
    \caption{\label{tab:table1} Capacitance setting for the one dimension scaling archtecture with grounded(differential) transmon couplers; the corresponding coupling strengths are shown in table.\ref{tab:table2}. The unit of capacitance is fF.}
    \begin{ruledtabular}
    \begin{tabular}{cccccc}
        identifier&$C_{t1}(C_T)$ &$C_{t2}$&$C_{f1}$
     &$C_{f2}$ &$C_c$ \\
    \hline
    1&  70 & $-$ & 6 &6  & 6 \\
    2&  50 & $-$ & 6 &6  & 6 \\
    3&  70 & $-$ & 6 &6  & 12 \\
    4&  70 & $-$ & 10 &6  & 6 \\
    5&  70 & $-$ & 6 &8  & 6 \\
    6&  20 &60& 6 &6  & 6 \\
    7&  40 &50& 6 &6  & 6 \\
    8&  80 &30& 6 &6  & 6 \\
    9&  20 &60& 6 &6  & 12 \\
    10&  40 &50& 6 &6  & 12 \\
    11& 80 &30& 6 &6  & 12 \\
    12& 80 &30& 10 &6  & 12 \\
    13& 60 &30& 10 &6  & 12 \\
    \end{tabular}
    \end{ruledtabular}
    \end{table}

\begin{table*}
    \caption{\label{tab:table2}Calculated Hamiltonian parameters of the one dimension scaling case corresponding to the set of capacitance of the same identifier in table.~\ref{tab:table1}.}
    \begin{ruledtabular}
    \begin{tabular}{ccccccccccc}
        identifier&$Ec_{1}($GHz$)$ &$Ec_{2}($GHz$)$&$Ec_{tc}($GHz$)$&$J_{12}($MHz$)$ 
        &$J_{1c}($MHz$)$&$J_{2c}($MHz$)$&$J_{13}($MHz$)$&$J_{1c2}($MHz$)$&$J_{cc}($MHz$)$&$J_{c1c3}($MHz$)$  \\
    \hline
    1&  1.942& 1.641& 0.250& -98.94& -399.6& 494.7& -0.957& -3.868& 19.34& 0.197 \\
    2&  1.945& 1.651& 0.337& -133.0& -539.0& 665.0& -1.734& -7.032& 35.16& 0.491 \\
    3&  1.859& 1.363& 0.239& -214.5& -547.1& 750.8& -4.222& -10.78& 37.74& 0.799 \\
    4&  1.598& 1.402& 0.246& -79.70& -374.2& 419.8& -0.490& -2.30& 12.11& 0.079 \\
    5&  1.617& 1.402& 0.249& -70.42& -332.3& 422.5& -0.776& -3.665& 21.99& 0.255 \\
    6&  1.951& 1.668& 0.269& -133.8& -215.8& -263.2& -1.763& -2.845& -5.595& -0.068 \\
    7&  1.943& 1.645& 0.269& -60.89& -215.5& -265.5& -0.363& -1.285& -5.603& -0.032 \\
    8&  1.940& 1.632& 0.269& -19.15& -215.3& -266.8& -0.036& -0.402& -5.607& -0.010 \\
    9&  1.876& 1.425& 0.266& -277.1& -306.2& -406.5& -7.129& -7.896& -11.58& -0.268 \\
    10&  1.863& 1.375& 0.266& -130.9& -305.2& -413.6& -1.577& -3.679& -11.63& -0.134\\
    11&  1.854& 1.344& 0.266& -42.07& -304.5& -417.9& -0.162& -1.173& -11.65& -0.044 \\
    12&  1.515& 1.180& 0.264& -37.21& -307.6& -368.0& -0.102& -0.839& -8.302& -0.022 \\
    13&  1.519& 1.191& 0.305& -55.90& -356.3& -424.9& -0.230& -1.464& -11.13& -0.045 \\
    \end{tabular}
    \end{ruledtabular}
    \end{table*}
\subsection{2D lattice}
\begin{figure*}[t!]
    \includegraphics[width=0.9\linewidth,height=5cm,keepaspectratio]{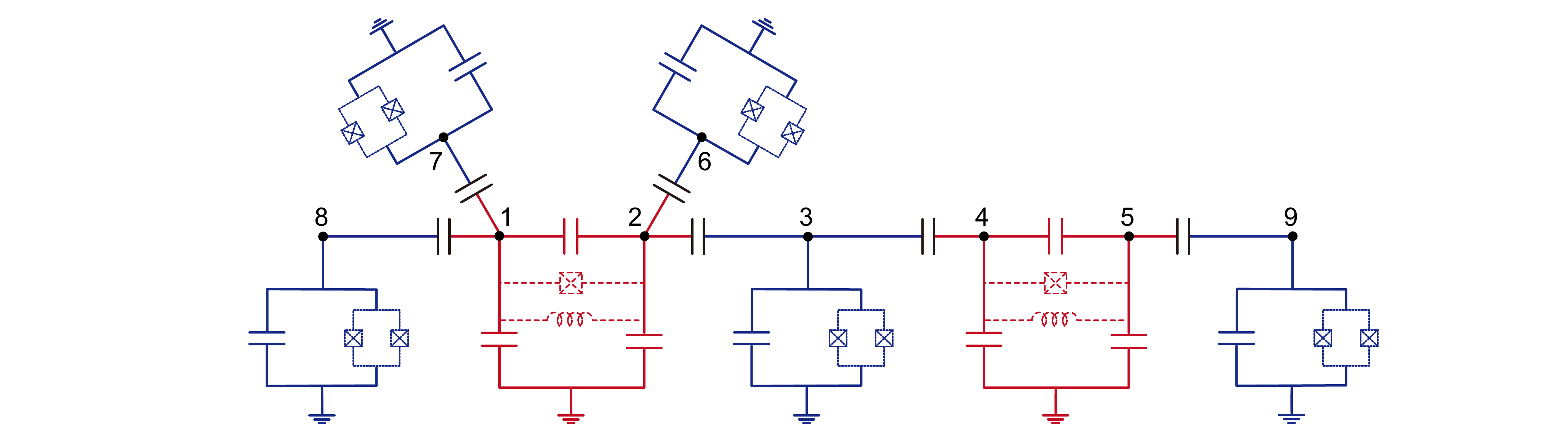}
    \caption{\label{fig:circuit3}Circuit model of FTF circuit of two-dimensional scaling with grounded transmons.}
    \end{figure*}
 Here we consider a square-lattice topology. A  fluxonium qubit (data qubit) is capacitively coupled to four transmons (couplers), as shown in Fig.~\ref{fig:circuit3}. Again, we neglect the small and qualitatively unimportant capacitance in the analysis. Elements of $\tilde{C}^{-1}$ for
this circuit are given by 
\begin{widetext}
    \begin{equation}\label{eqa5}
        \begin{aligned}
        \tilde{C}^{-1}[2,2]&\approx \frac{ 2C_c^2C_f+ C_{f1}^2(C_f+C_{f2})+ C_c(3C_f-C_f C_{f2}-2C_{f2}^2) }{2C_c(C_c C_f+2C_f^2-C_{f2})}\approx\frac{1}{\frac{5}{2}C_s}, \\
        \tilde{C}^{-1}[2,3]&\approx -\frac{C_c}{C_T(2C_c+C_f+C_{f2})}\approx-\frac{1}{5}\frac{1}{C_T}\frac{C_c}{C_s}, \\
        \tilde{C}^{-1}[2,5]&\approx -\frac{C_c^2 C_{f1}}{C_T[2C_c^2 C_f+C_{f1}(C_f+C_{f2})^2+C_c(3C_f^2+C_f C_{f2}-2C_{f2}^2)]}\approx -\frac{1}{25}\frac{1}{C_T}\frac{C_c}{C_s}\frac{C_c}{C_s}\\
        \tilde{C}^{-1}[3,6]&\approx \frac{2C_c^3 C_f}{C_T^2(4C_c^2 + 4C_c C_f+C_f^2-C_{f2}^2)}
        \approx\frac{4C_s}{15C_T^2}, \\
        \tilde{C}^{-1}[3,8]&\approx \frac{C_c^2 C_{f2}}{C_T^2(4C_c^2 + 4C_c C_f+C_f^2-C_{f2}^2)}
        \approx\frac{C_s}{15C_T^2}, \\
        \tilde{C}^{-1}[8,9]&\approx \frac{C_c^4 C_{f2}^2}
        {C_T^3[4C_c^4 + 12C_c^3C_f +C_c^2(13C_f^2-5C_{f2}^2)+(C_f^2-C_{f2}^2)^2+6C_c(C_f^3-C_f C_{f2}^2)]}\approx \frac{1}{120}\frac{C_c^2}{C_T^3}, \\
        \tilde{C}^{-1}[6,9]&\approx \frac{C_c^4 C_{f2}(C_f+2C_c)}
        {C_T^3[4C_c^4 + 12C_c^3C_f +C_c^2(13C_f^2-5C_{f2}^2)+(C_f^2-C_{f2}^2)^2+6C_c(C_f^3-C_f C_{f2}^2)]}\approx \frac{1}{30}\frac{C_c^2}{C_T^3}; \\
        \end{aligned}
    \end{equation}
    \end{widetext}
where the index $2,5$ are the fluxonium qubit modes and $3,6,7,8$ are the coupler modes, corresponding to Fig.~\ref{fig:circuit3}. It is obvious that with two more transmon couplers coupled to the fluxonium,  the effective capacitance further increases, while the modification of the qubit-coupler coupling strength and the qubit-qubit coupling strength can be neglected compared with the one dimension scaling case. Furthermore, the last two lines of Eq.~\ref{eqa5} imply that the coupling strength between the couplers on the same antenna is 4 times larger than that between the couplers attached to the different antennas, which makes this unwanted coupling the most deleterious one.
\begin{table*}
    \caption{\label{tab:table3}Calculated Hamiltonian parameters of the two dimension extension case corresponding to the set of capacitance of the same identifier in the table.~\ref{tab:table1}. Note that $J_{cc}$ is the coupling strength between two couplers on the different antennas, and $J^\prime_{cc}$ is that of couplers on the same antenna.}
    \begin{ruledtabular}
    \begin{tabular}{ccccccccccc}
        &$Ec_{1}($GHz$)$ &$Ec_{2}($GHz$)$&$Ec_{tc}($GHz$)$&$J_{12}($MHz$)$ 
        &$J_{1c}($MHz$)$&$J_{2c}($MHz$)$&$J_{cc}($MHz$)$&$J^\prime_{cc}($MHz$)$&$J_{c1c3}($MHz$)$&$J^\prime_{c1c3}($MHz$)$  \\
    \hline
    1&  1.329& 1.641& 0.247& -98.91& -399.8& 489.6& 11.11& 42.67& 0.113 &0.434 \\
    2&  1.345& 1.652& 0.333& -133.0& -539.5& 655.9& 20.75 & 78.56& 0.290 &1.096 \\
    3&  1.0& 1.859& 0.233& -157.5& -551.1& 727.1& 18.18& 98.42 & 0.385 &2.083\\
    4&  1.168& 1.598& 0.245& -66.37& -349.6& 372.4& 7.865& 35.46& 0.051 &0.230 \\
    5&  1.169& 1.617& 0.247& -58.63& -351.8& 329.8& 13.03& 40.80& 0.151 &0.472 \\
    \end{tabular}
    \end{ruledtabular}
    \end{table*}

\section{Perturbation theory of the static $ZZ$ interaction\label{ptzz}}
The static $ZZ$ interaction of FTF system can be understood by perturbation theory, where the second order correction to the energy of the state $\ket{m}$ is (here $m$ labels the eigenstates of the FTF Hamiltonian \eqref{eq:ftf1})
\begin{equation}
    E_m^{(2)}=\sum_{n_1}\frac{|V_{mn_1}|^2}{E^{(0)}_{mn_1}},
\end{equation}
where $n_1$ denotes any intermediate state different from{that is not} $\ket{m}$, $V_{j,k}=\bra{j}H\ket{k}$ represents the Hamiltonian matrix element between the bare states $\ket{j}$ and $\ket{k}$, while $E^{(0)}_{ij}=E^{(0)}_i-E^{(0)}_j$ are the corresponding energy gaps. The third order corrections are derived from the coupler-assisted interactions between the qubits 
\begin{equation}
    E_m^{(3)}=\sum_{n_1n_2}\frac{V_{mn_1}V_{n_1n_2}V_{n_2m}}{E^{(0)}_{mn_1}E^{(0)}_{mn_2}}.
\end{equation}
The fourth order corrections are contributed from both the direct interactions and the indirect interactions 
\begin{equation}
    E_m^{(4)}=\sum_{n_1n_2n_3}\frac{V_{mn_1}V_{n_1n_2}V_{n_2n_3}V_{n_3m}}{E^{(0)}_{mn_1}E^{(0)}_{mn_2}E^{(0)}_{mn_3}}-E^{(2)}_m\sum_{n_1}\left|\frac{V_{mn_1}}{E^{(0)}_{mn_1}}\right|^2.
\end{equation}
The corresponding static $ZZ$ interaction $\zeta=\zeta^{(2)}+\zeta^{(3)}+\zeta^{(4)}$ can be efficiently simulated by perturbation theory up to the fourth order, where the $k$-th order static $ZZ$ is of the form 
\begin{equation}
    \zeta^{(k)}=E_{11}^{(k)}+E_{00}^{(k)}-E_{10}^{(k)}-E_{01}^{(k)}.
\end{equation}
\section{Adiabatic CZ gate}
In this section we discuss the adiabatic CZ gate simulations in detail. All the codes are shared on Github \cite{code_github}.
In our simulation, we acquire the system Hamiltonian using Scqubits \cite{scqubits1,scqubits2}, and calculate the temporal evolution by Qutip \cite{QuTip}. For the fluxonium qubits, we obtain the bare states through diagonalization in the harmonic-oscillator basis with a truncation level of 150. For the transmons, we perform the diagonalization in the charge basis representation with a charge-number cutoff of 50. The state fidelity is calculated by standard definition $F(\hat{\sigma},\hat{\rho})=tr\sqrt{\hat{\rho}^{\frac{1}{2}}\hat{\sigma}\hat{\rho}^{\frac{1}{2}}}$, where $\hat{\sigma}$ is the ideal final state density matrix obtained by applying the ideal gate unitary $\hat{U}_g$ to the initial state and $\hat{\rho}$ is the simulated density matrix obtained by truncating the computational space (extended by $\ket{\widetilde{000}}$, $\ket{\widetilde{001}}$, $\ket{\widetilde{100}}$, and $\ket{\widetilde{101}}$) from the full Hilbert space used in simulation. We then estimate the gate fidelity by averaging the state fidelities over 36 initial two-qubit states generated from the set of six initial single-qubit states $\{\ket{0},\ket{1},\frac{1}{\sqrt{2}}(\ket{0}\pm\ket{1}),\frac{1}{\sqrt{2}}(\ket{0}\pm i\ket{1})\}$. 
\subsection{Optimization for coherent error}
As described in main text, the optimized pulse is composed of the constant-leakage-rate edge and the residual-leakage-suppressed flat duration. The edge profile is solved based on functions $D_{ij}(\varphi_{\rm ext,c})$ which are obtained through interpolations of the numerically calculated discrete values. Furthermore, although this optimized edge shape gives the maximum level of adiabaticity, small leakage error is still inevitable. To address this residual leakage problem, we develop a method to optimize the flat duration to make the leaked populations oscillate back to the computational states. 

\begin{figure} 
    \includegraphics[width=0.9\linewidth,keepaspectratio]{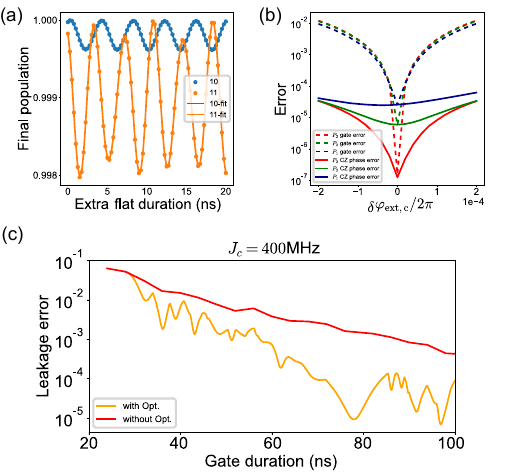}
    \caption{\label{fig:extral_flat}\textbf{Adiabatic CZ gate optimization.} (a) Final populations for the pulses with variable flat durations and fixed edge shape. In this simulation the pulses are acquired by adding extra flat durations into the $P_1$ pulse in main text. Dots in the figure represent the simulated results of final populations of the initial state of $\ket{\widetilde{100}}$ (orange) and $\ket{\widetilde{101}}$ (blue) at the end of the pulses, which are fitted by $p_1=A_1\sin(f_1 2\pi t+B_1)+C_1$and $p_2=A_2\sin(f_2 2\pi t+B_2) +A_3\sin(f_3 2\pi t+B_3)+C_2$. The fitting results yield $f_1=0.2506, f_2=0.2678, f_3=0.3840, $ and $ A_2/A_3=7.05$ represents the leakage population ratio of the two energy levels. (b) Simulated errors for pulses $P_1$, $P_2$, $P_3$ at various quasistatic flux deviations.(c) Simulation results of the leakage error of various gate pulses with the coupling strength $J_c=400{\rm MHz}$. The red line indicates the leakage errors without flat duration optimization, where the flat duration is settled to 0 ns. The orange line presents the minimal leakage errors with the optimization of the flat duration in our simulation.} 
    \end{figure}

\begin{figure*}[ht]
    \includegraphics[width=0.8\linewidth,keepaspectratio]{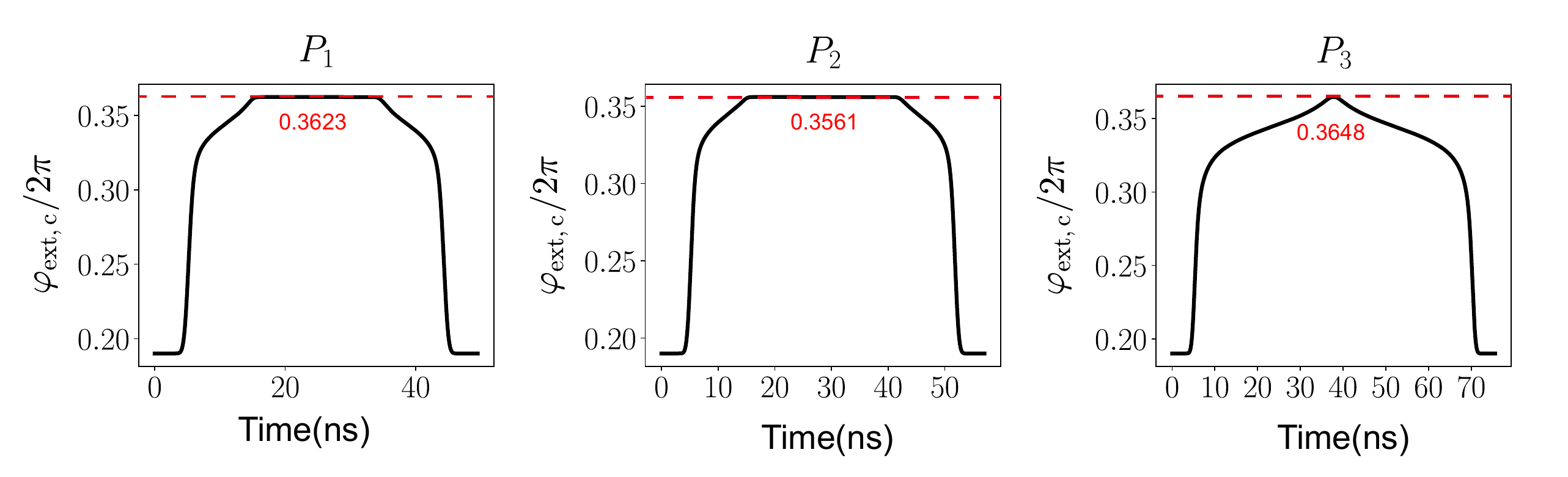}
    \caption{\label{fig:adpulses}\textbf{Pulse shapes for $P_1$, $P_2$, $P_3$.} } 
    \end{figure*}

This method can be understood by examining the gate dynamics. First, during the rising edge process, the population inevitably leaks to adjacent coupled energy levels in small amounts. After that, during the flat duration, the population in each energy level no longer changes; only the phase accumulates at different rates. This is equivalent to adjusting the initial phase for the Landau-Zener transitions \cite{196563,Zener} that occur during the dropping edge process. Thus, one can accumulate the appropriate phase in this way, ultimately ensuring that the leakage population returns to the computational states as much as possible by the end of the gate. As shown in Fig.~\ref{fig:extral_flat}(a), the final populations are oscillating with the pulses of various flat durations and fixed edge shape.
The oscillation frequency of the state $\ket{10}$ is $251$ MHz, corresponding to the frequency detuning between the dressed state $\ket{\widetilde{100}}$ and $\ket{\widetilde{010}}$ at the flux at the flat duration ($\varphi_{\rm ext,c}=0.362$). The oscillation of $\ket{11}$ state population has two frequency components, corresponding to the frequency detunings between the dressed state $\ket{\widetilde{101}}$ with the two leakage states $\ket{\widetilde{110}}$ ($268$ MHz) and $\ket{\widetilde{020}}$ ($365$ MHz). A fine optimized pulse occurs when the leakages of both computational states $\ket{10}$ and $\ket{11}$ are minimal. In our simulation we sweep the flat duration with various fixed edge durations to optimize the gate pulse, to acquire the minimal leakage error pulse shape at different total gate duration. Pulse shapes of the optimized pulse $P_1$, $P_2$, and $P_3$ are shown in Fig.~\ref{fig:adpulses}.
\subsection{Incoherent error simulation}\label{ap:ies}
We model the environment-induced dissipation as Markovian process, and analyze the effect on gate fidelity by master equation
\begin{equation}
    \dot{\rho}=-\frac{i}{h}[\hat{H},\rho]+\sum_k[\hat{L}_k\hat{\rho}\hat{L}_k^\dag-\frac{1}{2}(\hat{L}_k^\dag\hat{L}_k\hat{\rho}+\hat{\rho}\hat{L}_k^\dag\hat{L}_k)],
\end{equation}
where $\hat{L}_k$ is the Lindblad operators. In the simulation of couplers' dissipation it takes the form
\begin{equation}
    \hat{L}=\hat{I}_1\otimes(\sqrt{\Gamma_{1,c}}\ket{0}_c\bra{1}+\sqrt{\Gamma_{2,c}}\ket{1}_c\bra{2})\otimes\hat{I}_2,
\end{equation}
where $\hat{I}_1$ and $\hat{I}_2$ are the identity operator of fluxonium 1 and 2 respectively. The dissipation rate is related with coupler's $T_{1,c}$ by $\Gamma_{1,c}=1/T_{1,c}$, and here we assume $\Gamma_{2,c}=2\Gamma_{1,c}$. Similarly, the Lindblad operators for fluxoniums' dissipation are
\begin{equation}
    \begin{aligned}
    \hat{L}'_1&=\sqrt{\Gamma_{1}}\ket{0}_1\bra{1}\otimes\hat{I}_c\otimes\hat{I}_2\\
    \hat{L}'_2&=\hat{I}_1 \otimes\hat{I}_c\otimes\sqrt{\Gamma_{2}}\ket{0}_2\bra{1},
    \end{aligned}
\end{equation}
where $\hat{I}_c$ is the identity operator of the coupler, and in the simulations we choose $\Gamma_1=\Gamma_2=1/T_{1,f}$ for simplicity.

For frequency-tunable transmon, flux noise is the main source of pure dephasing process. Here we use a quasi-static Gaussian noise model to simulate the induced error and relate it with $1/f$ noise power spectrum density of the form $S_{\varphi}(f) = A_{\varphi}^2/f$ by equation $\sigma^2_{\varphi}=\int_{f_\mathrm{IR}}^{f_\mathrm{UV}}S_{\varphi}(f')df'$, where $\sigma_{\varphi}=\sqrt{\langle\delta\varphi_{\rm ext,c}\rangle}$ is the standard deviation of the Gaussian distribution of the flux fluctuation, and $f_\mathrm{IR}$, $f_\mathrm{UV}$ is the infrared and ultraviolet cut-off frequency respectively. In the simulation we choose $f_\mathrm{IR}=10^{-6}$ Hz and $f_\mathrm{UV}=1$ GHz. We simulate the gate fidelity with the pulse $P_1$, $P_2$ and $P_3$ in the main text at various flux deviations, and then use the results to calculate the average gate fidelity $\langle F_g \rangle$ of the Gaussian distribution with various standard deviation.
In Fig.~\ref{fig:extral_flat}(b), the CZ gate fidelities (dashed lines) are calculated by the ideal CZ gate unitary $U_g=(U_{\phi_1}\otimes U_{\phi_2})U_{CZ}$, where
\begin{equation}
    \begin{aligned}
        U_{CZ}&=\begin{pmatrix}
            1    & 0  &0 & 0 \\ 
            0    & 1  &0 & 0 \\ 
            0    & 0  &1 & 0 \\ 
            0    & 0  &0 & e^{i\pi} \\ 
        \end{pmatrix},\\
        U_{\phi_i}&=\begin{pmatrix}
            1    & 0   \\ 
            0    & e^{i\phi_i}  \\ 
        \end{pmatrix}.
    \end{aligned}
\end{equation}
The single-qubit phases $\phi_1$ and $\phi_2$ are fixed and extracted from the case of zero flux deviation. To further explore the error type, we also calculate the CZ phase error (with optimizing $\phi_1$ and $\phi_2$), and the leakage error (with optimizing $\phi_1$, $\phi_2$ and also CZ phase modification $\delta\theta$), where the unitary of CZ  gate of a phase modification $\delta\theta$ is
\begin{equation}
        U_{CZ,\delta\theta}=\begin{pmatrix}
            1    & 0  &0 & 0 \\ 
            0    & 1  &0 & 0 \\ 
            0    & 0  &1 & 0 \\ 
            0    & 0  &0 & e^{i(\pi+\delta\theta)} \\ 
        \end{pmatrix}.
\end{equation}
The results reveal that the leakage error is negligible (not plotted), while the main error comes from the single-qubit phase deviations (Fig.~\ref{fig:extral_flat}(b)). 

\section{Microwave-activated CZ gate\label{appendix:map_gate}}
In the simulation of the microwave-active CZ gate, we choose the Hamiltonian parameters as: $E_{C,1(2)}=1.2$ GHz, $E_{J,1(2)}=6.1(6)$ GHz, $E_{L,1(2)}=0.17(0.15)$ GHz, $E_{C,c}=300$ MHz, $E_{J,c}=25$ GHz, $J_{C,1(2)}=450$ MHz, and $J_{12}=120$ MHz. The $0-1$ transition frequencies of the fluxoniums are 3.214 GHz and 2.848 GHz, respectively. The target transition frequency is \(5.948\) GHz at \(\varphi_{\rm ext,c}/2\pi = 0.21\). These frequencies help in determining the appropriate Hilbert space dimension for the FTF system in numerical calculations. Our approach is to include all the possible energy levels that might be excited during the gate process. To ensure this, we set a safe threshold for the highest energy level at \(\max_{ij,ts}\{E_{ij,ts}\} + E_T \approx 12\) GHz. In this paper, we retain 10 bare state levels for each fluxonium and four levels for the transmon. The frequencies of the highest energy levels for the fluxoniums are 13.839 GHz and 13.398 GHz, while for the transmon, it is 14.967 GHz at \(\varphi_{\rm ext,c}/2\pi = 0.32\). Consequently, the total dimension of the static Hamiltonian is 400. For the simulation of the driving dynamics, we truncate to the lowest 50 energy levels, among which the highest frequency being 14.326 GHz at \(\varphi_{\rm ext,c}/2\pi = 0.21\).
\begin{figure}[ht]    \includegraphics[width=0.9\linewidth,keepaspectratio]{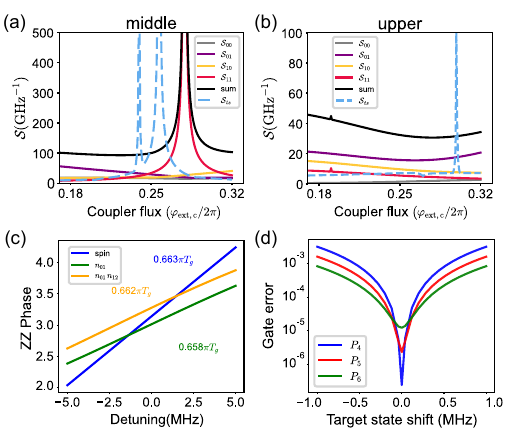}
    \caption{\label{fig:mw_ap1}\textbf{Microwave-activated CZ gate simulation.} (a)(b) $\mathcal{S}$ parameters of middle and upper state in the three-level subspace. (c) Simulation results of the phase tuning rate. (d) Gate fidelities with target state frequency shifts caused by stray couplings.} 
    \end{figure}

In the main text, we select the lower energy level within the three-level subspace as the target state. However, it is also possible to choose the middle or upper energy levels as the target. The values of \(\mathcal{S}_{ts}\) remain low in regions away from on-resonance peaks. Consequently, when targeting the lower state, the primary leakage error arises from the population escaping from the target state, leading to the leakage of the \(\ket{00}\) state. Conversely, when targeting the middle or upper states, the main leakage error is due to leakage from the \(\ket{01}\) state as shown in Fig.~\ref{fig:mw_ap1}(a) and (b).

When calculating the phase tunability for our Gaussian pulse, we first use the spin model, resulting a rate of $0.663\pi T_g$. Then we use the full Hamiltonian containing 50 energy levels, but only include the charge matrix element $n_{01}$ by setting $n_{12}$, $n_{23}$ and $n_{34}$ equal to zero. Next we further include both $n_{01}$ and $n_{12}$. The corresponding results are shown in Fig.~\ref{fig:mw_ap1}(c). 

Due to the spectator effect, spectator coupling can shift the target state frequency, introducing leakage and phase deviations in the CZ gate. At a coupler flux of $\varphi_\mathrm{ext,c}=0.22\times 2\pi$, our analysis reveals that a 10 MHz stray coupling induces a modest 0.15 MHz shift in the coupler frequency. As shown in Fig.~\ref{fig:mw_ap1}(d), the microwave-activated gates retain fidelities above $99.99\%$ despite this perturbation—demonstrating the robustness of our implementation.
\begin{figure}[h!]
    \includegraphics[width=0.9\linewidth,keepaspectratio]{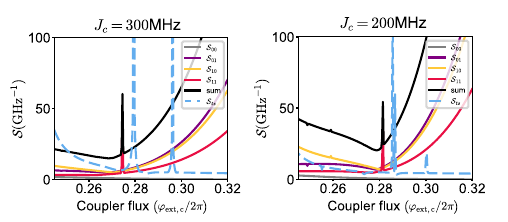}
    \caption{\label{fig:mw_ap2}\textbf{$\mathcal{S}$ parameter with $J_C = 300$ MHz and $J_C = 200$ MHz.} } 
    \end{figure}

Additionally, we study $\mathcal{S}$ with a smaller qubit-coupler coupling ($J_C = 300$ MHz and $J_C = 200$ MHz while maintaining the ratio $J_C/J_{12} = 5$). The results indicate that for $J_C = 300$ MHz, there is still an optimized range, but it corresponds to a narrower frequency span of approximately 200 MHz. In contrast, for $J_C = 200$ MHz, the values of $S_{\text{sum}}$ increase, suggesting that the leakage error may not be effectively suppressed with our approach.

\providecommand{\noopsort}[1]{}\providecommand{\singleletter}[1]{#1}%

\end{document}